\documentclass[pra,aps,10pt,superscriptaddress,floatfix,twocolumn]{revtex4-2}
\usepackage{graphicx,color,xcolor}
\usepackage[nice]{nicefrac}	
\usepackage{amsmath,amssymb,bm}
\usepackage{amsmath}
\usepackage{braket}
\usepackage[rightcaption]{sidecap}
\usepackage{lineno}
\usepackage{etoolbox}
\usepackage{enumitem}
\setlist[itemize]{align=parleft,left=0pt..1em}
\makeatletter
\patchcmd{\frontmatter@abstract@produce}
  {\vskip200\p@\@plus1fil
   \penalty-200\relax
   \vskip-200\p@\@plus-1fil}
  {}
  {}
  {}
\makeatother

\usepackage[plainpages=false,pdfpagelabels,colorlinks=true,linkcolor=red,urlcolor=blue,citecolor=blue,pdftitle={},pdfauthor={},pdfdisplaydoctitle=true,pdfduplex=DuplexFlipLongEdge]{hyperref}

\definecolor{darkred}{rgb}{0.90,0.2,0.2}
\definecolor{darkgreen}{rgb}{0,0.60,.2}
\definecolor{darkblue}{rgb}{0.1,0.3,1}
\definecolor{gray}{cmyk}{0,0,0,0.25}
\definecolor{orange}{cmyk}{0,0.6,0.8,0}

\usepackage{graphicx} 

\begin{document}
\title{Timescales and necessary conditions for hydrodynamization in \\ one-dimensional Bose gases}

\author{Yicheng Zhang}
\affiliation{Homer L. Dodge Department of Physics and Astronomy,\\
The University of Oklahoma, Norman, Oklahoma 73019, USA}
\affiliation{Center for Quantum Research and Technology, The University of Oklahoma, Norman, Oklahoma 73019, USA}
\author{Yuan Le}
\affiliation{Department of Physics, The Pennsylvania State University, University Park, Pennsylvania 16802, USA}
\author{David S. Weiss}
\affiliation{Department of Physics, The Pennsylvania State University, University Park, Pennsylvania 16802, USA}
\author{Marcos Rigol}
\affiliation{Department of Physics, The Pennsylvania State University, University Park, Pennsylvania 16802, USA}

\date{\today}

\begin{abstract}
We study the quantum evolution of one-dimensional Bose gases immediately after several variants of high-energy quenches, both theoretically and experimentally. Using the advantages conveyed by the relative simplicity of these nearly integrable many-body systems, we are able to differentiate the behaviors of two distinct but often temporally overlapping processes, hydrodynamization and local prethermalization. We show that the hydrodynamization epoch is itself characterized by two independent timescales, an oscillation period and an observable-dependent damping time. We also show how the existence of a hydrodynamization epoch depends on the exact nature of the high-energy quench. There is a universal character to our findings, which can be applied to the short-time behavior of any interacting many-body quantum system after a sudden high-energy quench. We specifically discuss its potential relevance to heavy-ion collisions.
\end{abstract}

\maketitle

\section{Introduction}

The term ``hydrodynamization'' was introduced to describe the notably short period of time between the start of a relativistic heavy-ion collision and when hydrodynamics describes its time evolution~\cite{prethermalization, Spalinksy_2015, Florkowski_2018, busza_18, schlichting_2019, Schenke_2021, RevModPhys_2021}. The observed hydrodynamization time is shorter than the expected local thermalization time, which is the time required for the system to reach thermal equilibrium within a hydrodynamic cell. Based on experimental results with nearly integrable gases of ultracold atoms in one dimension and the associated theory, we argued in Ref.~\cite{le2023observation} that immediately after a high-energy quench the time evolution of any many-body quantum system can be qualitatively understood in a universal way. There is an initial epoch, which by analogy to the relativistic heavy-ion usage we called ``hydrodynamization," during which energy is redistributed across distant momentum modes and short-distance density fluctuations damp out. The second, partially temporally overlapping epoch is local prethermalization, which is analogous to thermalization in far-from-integrable systems and is characterized by a redistribution of energy among nearby momentum modes. Hydrodynamization and local prethermalization are quantum mechanical phenomena, yet we showed in the experiments in Ref.~\cite{le2023observation} that generalized hydrodynamics (a classical description)~\cite{castro2016emergent, bertini2016transport, doyon_20, Alba_2021, Bastianello_2022} can be used to describe the system before local prethermalization is complete as long as the hydrodynamization epoch is over.

In this paper, again using nearly integrable one-dimensional (1D) gases as a model, we significantly broaden our understanding of hydrodynamization. In addition to the universal hydrodynamization coherence time that we identified in Ref.~\cite{le2023observation}, we identify a new observable-dependent hydrodynamization damping time that sets the lower limit for the application of hydrodynamic descriptions. We show, in both theory and experiments, that the existence of a distinct hydrodynamization epoch requires that there be a multimodal energy distribution after the quench. We further show that, although the extent of temporal overlap of hydrodynamization damping (when it exists) and local prethermalization depends on the system and the observable being considered, the two timescales scale with initial state parameters in the same way.  We are able to apply these findings to a quantitative consideration of relativistic heavy-ion collisions. Our estimates of the shortest time after which hydrodynamics can describe the dynamics in those systems are in line with the observations of those collisions.

To see how a nearly integrable system can give insights into nonintegrable system dynamics, it is useful to start with a very high level view of dynamics in quantum systems. The time-evolving wave function $\Psi(\textbf{q}, t)$ describing an isolated quantum system can be written as
\begin{equation}\label{eq:Bohr}
\Psi(\textbf{q},t)=\sum_{j}\, c_j\, \psi_j(\textbf{q})\, e^{-iE_j t/\hbar},
\end{equation}
where $\textbf{q}$ includes the positions of all the particles, $\psi_j(\textbf{q})$ are the energy eigenfunctions with eigenenergies $E_j$, and the (time independent) $c_j$ are the overlaps of the initial state with $\psi_j(\textbf{q})$. Unitary dynamics is the result of the relative evolution of the phases $E_j t/\hbar$ between the energy eigenfunctions. Equation~\eqref{eq:Bohr} contains all the system's dynamical information but it is not useful for large, generic interacting quantum systems, where none of its elements can be calculated. Without explicit calculations, the equation generally provides no way to extract the timescales of processes such as hydrodynamization and local prethermalization. 

Nearly integrable many-body systems, however, are special in this regard. They can be characterized in terms of very long lived quasiparticles that arise from interactions among particles. The momenta of these quasiparticles are known as rapidities $\theta$, the distribution of which fully characterizes equilibrium states in the integrable limit~\cite{korepin_bogoliubov_book_93}. Rapidity distributions, $f(\theta)$, can be calculated in experimentally relevant models and can be measured in 1D gases~\cite{wilson_malvania_20, malvania_zhang_21, li_zhang_23, yang2023phantom}. Consideration of the relative evolution of the phases associated with quasiparticle states allows one to identify important out-of-equilibrium features, including the timescales associated with hydrodynamization and local prethermalization. Notably, such features can be identified even when exact calculations are out of reach. We conjectured in Ref.~\cite{le2023observation} that, shortly after high-energy quenches, far-from-integrable quantum systems admit a similar description that incorporates interactions into quasiparticles, with the distinction that they are short-lived and likely hard to calculate. But since the hydrodynamization epoch is so short, quasiparticles are not expected to decay in that time. We therefore expect that the expanded understanding of hydrodynamization we develop here from consideration of nearly integrable 1D gases can be applied to generic interacting quantum systems, as we do here for relativistic heavy-ion collisions. 

The presentation is organized as follows. In Sec.~\ref{sec:TGbragg}, we introduce the experimental setup and review its theoretical modeling. We also present our overall understanding of the quantum dynamics that follow a single Bragg pulse acting on a Tonks-Girardeau gas. In Sec.~\ref{sec:3peak}, we theoretically study hydrodynamization and local prethermalization in a translationally invariant system with a trimodal rapidity distribution. Changing the width of the rapidity peaks and their separation allows us to probe, for the first time, the two timescales associated with hydrodynamization as well as the local prethermalization time for the momentum distribution. In Sec.~\ref{sec:2peak}, we consider the case of a symmetric bimodal rapidity distribution, which allows us to demonstrate the absence of hydrodynamization in high energy quenches that lack multimodal energy distributions. The latter conclusion is tested experimentally in Sec.~\ref{sec:qnc} using a quantum Newton's cradle~\cite{kinoshita2006quantum} initial state. The relevance of our findings to heavy-ion collisions is discussed in Sec.~\ref{sec:heavyion}. We summarize our results, and highlight open questions in Sec.~\ref{sec:summary}.

\section{Overview}\label{sec:TGbragg}

\subsection{Bragg pulse quench in 1D Bose gases,\\ experiment and theory}\label{sec:ooeg}

We previously studied hydrodynamization in 1D gases both experimentally and theoretically. Our 1D gas experiments are implemented with a blue-detuned two-dimensional (2D) lattice, which confines atoms in a bundle of tube-like traps, and an independent red-detuned light trap which confines them axially. The experiment allows for a range of dimensionless coupling strengths. Our theory was done in the hard-core limit for atoms in a single 1D trap with parameters that match those of the experimental setup. In both the experiment and the theory, we took the 1D gases initially in their ground states out of equilibrium via a Bragg pulse quench with a $\lambda$ wavelength standing wave, and then watched the subsequent evolution of the momentum distribution, $f(p)$. Appendix~\ref{app:exp-the-bra} reviews the details of our experimental and theoretical setups.

Our earlier work found qualitative agreement between the experiment and theory in the postquench evolution of momentum distributions, both with regard to hydrodynamization and local perthermalization (see also Appendix~\ref{app:nsbp}). Still, there were quantitative differences related to the difference between finite and infinite coupling strengths. For instance, the difference between the timescales associated with hydrodynamization and local prethermalization is about three times larger in the experiment than in theory, so these processes can be better resolved in the experiments. But our experiments come with several limitations: $\lambda$ is fixed; there is a limited range over which the initial average energy of the gases can be varied; there is density inhomogeneity within each tube; the Bragg pulses take some minimum time; and the Bragg pulses create nonzero higher order Bragg peaks.

In this paper, we decouple the theory from these experimental limitations, doing away with the trap by theoretically modeling homogeneous 1D gases and replacing the Bragg pulses with tailored initial rapidity distributions. In Appendix~\ref{app:initialstate}, we show that the calculated dynamics of the momentum distribution are not significantly affected by removing either the trap or the Bragg pulses. Our results therefore retain experimental relevance even with this simpler modeling. Our theory-centric approach allows us to vary $\lambda$ and the initial average energy of the gases at will, which is what allows us to distinguish two distinct timescales within the hydrodynamization epoch. We also avoid unwanted transitions to second order Bragg peaks, which allows us to unambiguously show what kind of momentum distributions are needed for hydrodynamization to occur. In Sec.~\ref{sec:qnc}, we illustrate this finding in our experimental system.

\begin{figure}[!t]
    \includegraphics[width=0.9\columnwidth]{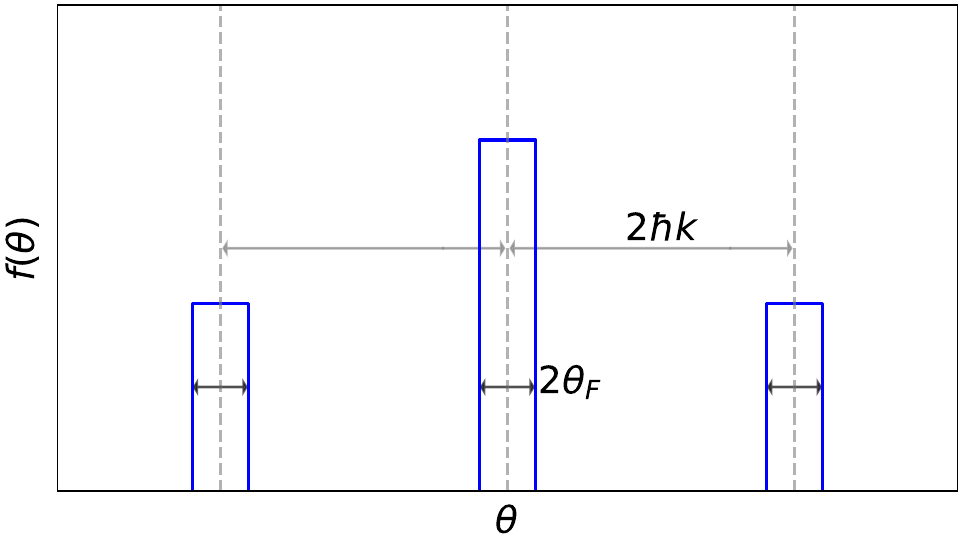}
    \vspace{-0.15cm}
    \caption{{\it Schematic trimodal rapidity distribution $f(\theta)$.} The rapidity peaks are $2\hbar k$ apart and have a width $2\theta_F$.}
    \label{fig:rapiditysketch}
\end{figure}

A qualitative discussion of the dynamics after a quench can be made with reference to a trimodal rapidity distribution, $f(\theta)$, like that shown in Fig.~\ref{fig:rapiditysketch}. The rapidity distribution is conserved, so evolution of the many-body state proceeds due to the different phase evolutions of the rapidities, which have associated energies of $(\hbar \theta)^2/2m$, similar to Eq.~\eqref{eq:Bohr} ($m$ is the quasi-particle mass). It also depends on the initial phase differences among the rapidities. The time evolution of the relative phases of different rapidities enters into $f(p)$ via a Bose-Fermi mapping (see Appendix~\ref{app:exp-the-bra}), so $f(p)$ changes in time despite $f(\theta)$ being constant. That is, although there is no redistribution of energy among the quasiparticles, there is a redistribution of energy among the particles.

\subsection{Hydrodynamization oscillations}\label{sec:hydro}

The largest relative energy difference between significantly populated quasiparticle states sets the fastest rate at which $f(p)$ can evolve. This is the ``hydrodynamization coherence'' time $T^c_{\rm hd}$. For the distribution in Fig.~\ref{fig:rapiditysketch}, $\Delta E\simeq (2\hbar k)^2/2m$ (where $k=2\pi/\lambda$ is the wave number), which gives a characteristic hydrodynamization oscillation frequency $\omega^c_{\rm hd}=\hbar (2k)^2/2m$ and a corresponding period 
\begin{equation}\label{eq:thd}
T^c_{\rm hd}=\frac{2\pi}{\omega^c_{\rm hd}}=\frac{\pi m}{\hbar k^2}. 
\end{equation}
In Ref.~\cite{le2023observation} we observed energy redistribution into and out of the modes near the Bragg peaks at a frequency $2\omega^c_{\rm hd}$. Theoretical calculations in the TG limit showed complementary behavior in position space; the $\lambda/2$-scale variations of density generated by the Bragg pulse oscillate at $\omega^c_{\rm hd}$. One can understand that temporal oscillations in momentum space are twice as fast as temporal oscillations in position space because momentum distribution changes depend on the absolute value of the departure of the density distribution from its average value.

\subsection{Hydrodynamization damping}\label{sec:hydrodamping}

While there is a typical hydrodynamization oscillation frequency associated with the energy set by the peak separation, the width of the peaks introduces a dispersion about that frequency that leads to damping of hydrodynamization oscillations in both position and momentum space, which was also observed in Ref.~\cite{le2023observation}. 

The bandwidth of energy differences between the center and side peaks, $W$, depends on the width of the sidepeaks, $2\theta_F$, and on the rapidity modes that are coupled. If the rapidities of the coupled modes differ by $\pm 2\hbar k$, as is the case of interest in this work, then $W=8\hbar k \theta_F/(2m)$, so the corresponding dephasing timescale is
\begin{equation}\label{eq:tdp}
    T_{\rm dp}=\frac{2\pi\hbar}{W}=\frac{\pi m}{2k\theta_F}.
\end{equation}
The specific way in which an observable evolves during hydrodynamization depends on the range, amplitudes, and relative phases of quasiparticle states that contribute to that observable. The hydrodynamization oscillations of any observable $\hat O$ will damp in a time $T^d_{\rm hd}(\hat O)$ that is proportional to $T_{\rm dp}$, with a proportionality constant that depends on the observable. 

One of the main goals of this work is to disentangle the two timescales $T^c_{\rm hd}$ and $T^d_{\rm hd}$, which we achieve by changing $\hbar k$ and $\theta_F$ independently in our model calculations. We will show here that $T^c_{\rm hd}$ and $T^d_{\rm hd}$ change with $\hbar k$ and $\theta_F$ as predicted by Eqs.~\eqref{eq:thd} and~\eqref{eq:tdp}, respectively, i.e., they can be very different if $2 \theta_F$ and $\hbar k$ are very different. It is only after the entire hydrodynamization epoch is over, that is, after hydrodynamization oscillations have damped out, that hydrodynamic descriptions are possible. Short-distance scale fluctuations, which are not captured in hydrodynamic descriptions, damp out only when the hydrodynamization oscillations damp out.

\subsection{Local prethermalization}\label{sec:locpreth}

Local prethermalization also results from the width ($2\theta_F$) of the rapidity peaks. Unlike hydrodynamization dephasing, local prethermalization is not generally associated with the damping of significant spatial fluctuations or the redistribution of amplitudes among distant momentum modes. Rather, for $f(p)$, local prethermalization describes the exchange of particles among nearby momentum modes, as was shown in Ref.~\cite{le2023observation}. Local prethermalization is the only thing that happens after a low-energy quench or after a high-energy quench in which the energy distribution of the quasiparticles is not multimodal. Like $T^d_{\rm hd}$ and for the same reason, we expect the local prethermalization time $T_{\rm lp}(\hat O)$ to be proportional to $T_{\rm dp}$ for all observables $\hat O$ [$T_{\rm lp}(\hat O)\propto T_{\rm dp}$] with observable-dependent coefficients that are in general different from those of $T^d_{\rm hd}(\hat O)$. Consistent with this expectation, and as expected on physical grounds, in Ref.~\cite{le2023observation} the prethermalization time for $f(p)$ was found to increase as $p$ decreases for values of $p$ within the central momentum peak.

While local prethermalization (thermalization) occurs in general after quantum quenches in near-integrable (far-from-integrable) systems, hydrodynamization only occurs after quantum quenches that introduce an energy density that is much greater than that of the ground state~\cite{le2023observation}. Another main goal of this work is to show that, for hydrodynamization to occur, in addition to the quenches being high-energy, the energy distribution of the quasi-particles after the quench must be multimodal.

\section{Three-peak states}\label{sec:3peak}

To flesh out the distinction between $T^c_{\rm hd}$ and $T^d_{\rm hd}$, and between $T^d_{\rm hd}(\hat O)$ and $T_{\rm lp}(\hat O)$, we use a particular wave function that gives rise to the distribution in Fig.~\ref{fig:rapiditysketch}, which we refer to as the three-peak state, 
\begin{equation}\label{eq:3peaks}
|\Psi_{\rm 3p}\rangle=\prod_{|\theta|<\theta_F}\!\! \left(-i A_{-1}\hat c^{\dagger}_{\theta-2\hbar k} + A_0 \hat c^{\dagger}_{\theta} -i A_{1}\hat c^{\dagger}_{\theta+2\hbar k} \right)|0\rangle\,,
\end{equation}
where we have chosen the phases of each ``replica'' of the initial Fermi sea (with Fermi momentum $\theta_F$) to match those of the Bessel functions in Eq.~\eqref{eq:braggwf}, making the superpositions similar to the experiments in Ref.~\cite{le2023observation} and in this work. Choosing other phases does not qualitatively change any of the scalings discussed in what follows. We select the weights of the peaks to be $A_{0}=2/\sqrt{6}$ and $A_{\pm 1}=1/\sqrt{6}$, which are different from the values $A_0\approx\sqrt{2}/2$ and $A_{\pm 1}\approx 1/2$ that we use in Appendix~\ref{app:initialstate} to model our experiment. As we will see, changing $A_0$ and $A_{\pm 1}$ does not qualitatively change the hydrodynamization and prethermalization phenomenology. For all the calculations in this section, we fix the length of our periodic 1D system to be $L=20\times2\pi/k_0$, where $k_0$ is the wave number from the experiments. To study the effect that changing the particle density (the Fermi momentum $\theta_F$) has on the different timescales, we change the particle number $N$. 

\begin{figure}[!t]
    \includegraphics[width=0.985\columnwidth]{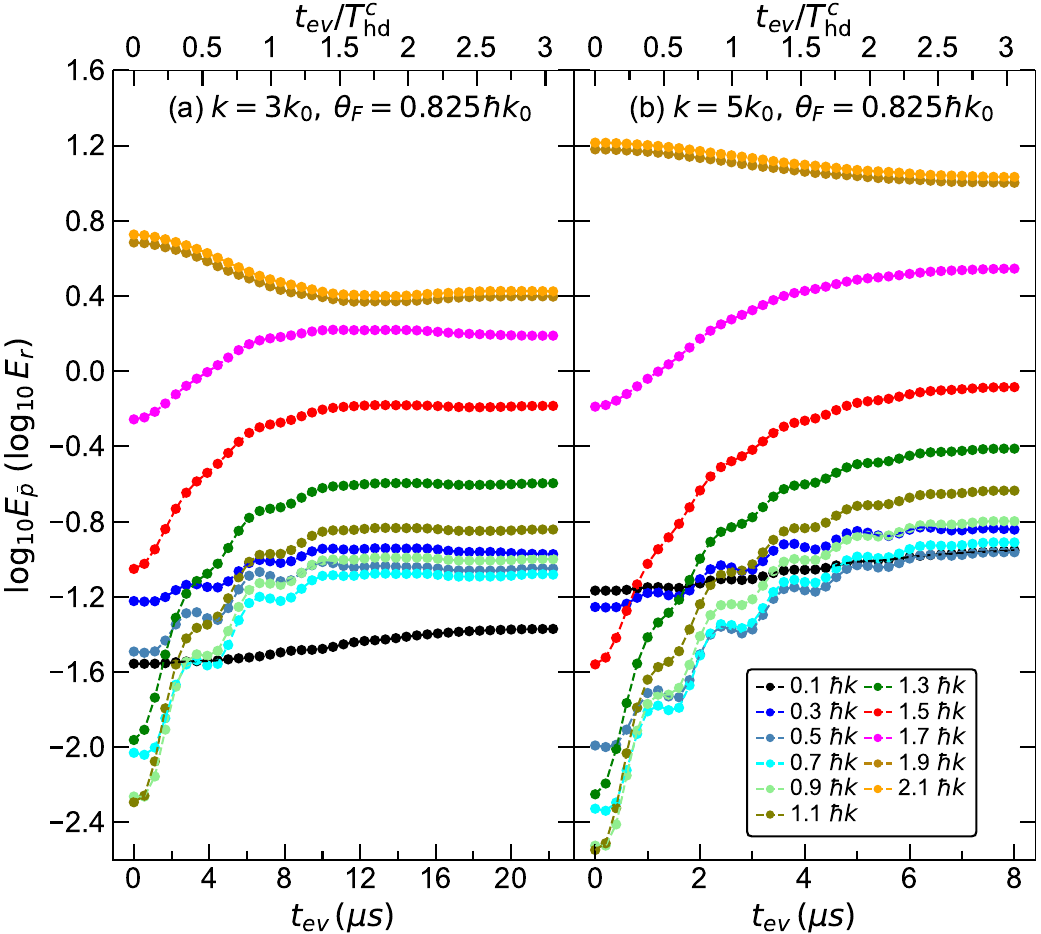}
    \vspace{-0.15cm}
    \caption{{\it Time evolution of $E_{\bar p}$ for three-peak states} [{\it see Eq.}~\eqref{eq:3peaks}] {\it with $A_{0}=2/\sqrt{6}$ and $A_{\pm 1}=1/\sqrt{6}$.} We show results for systems with $N=33$ ($\theta_F=0.825\hbar k_0$) for (a) $k=3k_0$ and (b) $k=5k_0$. Each curve is integrated within a 0.2 $\hbar k$ momentum interval. Note that the time axes differ in the two panels. The upper $x$ axis shows the dimensionless quantity $t_{\rm ev}/T^c_{\rm hd}$.}
    \label{fig:3peakek}
\end{figure}

In Fig.~\ref{fig:3peakek}, we plot the time evolution of $E_{\bar p}$ [the energy within each slice of momentum, see Eq.~\eqref{eq:barEp}] for two values of $k$, $k=3k_0$ [Fig.~\ref{fig:3peakek}(a)] and $k=5k_0$ [Fig.~\ref{fig:3peakek}(b)] and the same $N=33$. When plotted in terms of the dimensionless time $t_{\rm ev}/T^c_{\rm hd}$ (see the upper $x$ axes labels), the hydrodynamization oscillation frequency is the same for all $p$ and the same in both panels. That is, the hydrodynamization oscillation period is proportional to $k^{-2}$, as expected from Eq.~\eqref{eq:thd}.  Hydrodynamization damping is characterized by how long the hydrodynamization oscillations last. Comparing the curves for $E_{\bar p}$ with the same relative $\bar p$ in the two panels, we see that the damping times scale as $k^{-1}$ for fixed $\theta_F$, as expected from Eq.~\eqref{eq:tdp}. We also see that the approach to steady state values, a consequence of local prethermalization, scales as $k^{-1}$. The difference in the behavior of $E_{\bar p}(t_{\rm ev})$ for different values of $\bar p$ within each panel illustrates how hydrodynamization damping and local prethermalization play out differently for different observables, a manifestation of the fact that each observable (in this case each $E_{\bar p}$) has a unique dependence on the rapidity distribution.

\begin{figure}[!t]
    \includegraphics[width=0.985\columnwidth]{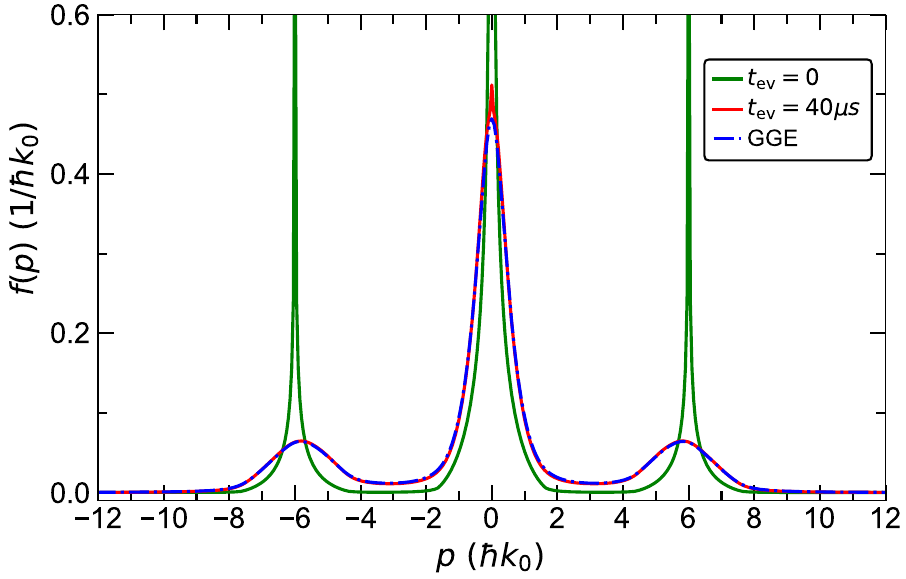}
    \vspace{-0.15cm}
    \caption{{\it Evolution of the momentum distribution $f(p)$ for a three-peak state.} The solid lines show $f(p)$ at $t_{\rm ev}=0$ (green) and 40 $\mu$s (red) for the same initial state as in Fig.~\ref{fig:3peakek}(a). The blue dashed line shows the GGE prediction.}
    \label{fig:3peakfp}
\end{figure}

Next, we study the prethermalization of the momentum distribution in detail. As shown in Refs.~\cite{rigol2007relaxation, rigol2007relaxation2, gramsch_12}, the integrability of the Tonks-Girardeau gas implies that the momentum distribution after equilibration is not thermal. Instead, it is described by a generalized Gibbs ensemble (GGE) that is fully determined by the conserved rapidity distribution of the underlying quasiparticles. Similar results have been obtained in other paradigmatic integrable models that map onto quadratic ones, such as the Luttinger model~\cite{cazalilla_06} and the transverse-field Ising model in one dimension~\cite{vidmar2016generalized, calabrese_11, fagotti_13} (see Refs.~\cite{ilievski_denardis_15, calabrese_essler_review_16} for other examples). In Fig.~\ref{fig:3peakfp}, we show $f(p)$ for the state studied in Fig.~\ref{fig:3peakek}(a) at $t_{\rm ev}=0$ and 40 $\mu$s, along with the GGE prediction. Except for the occupation of $p=0$, which requires a longer time to equilibrate, $f(p)$ at $t_{\rm ev}=40$ $\mu$s is indistinguishable from the GGE prediction. 

\begin{figure}[!t]
    \includegraphics[width=0.985\columnwidth]{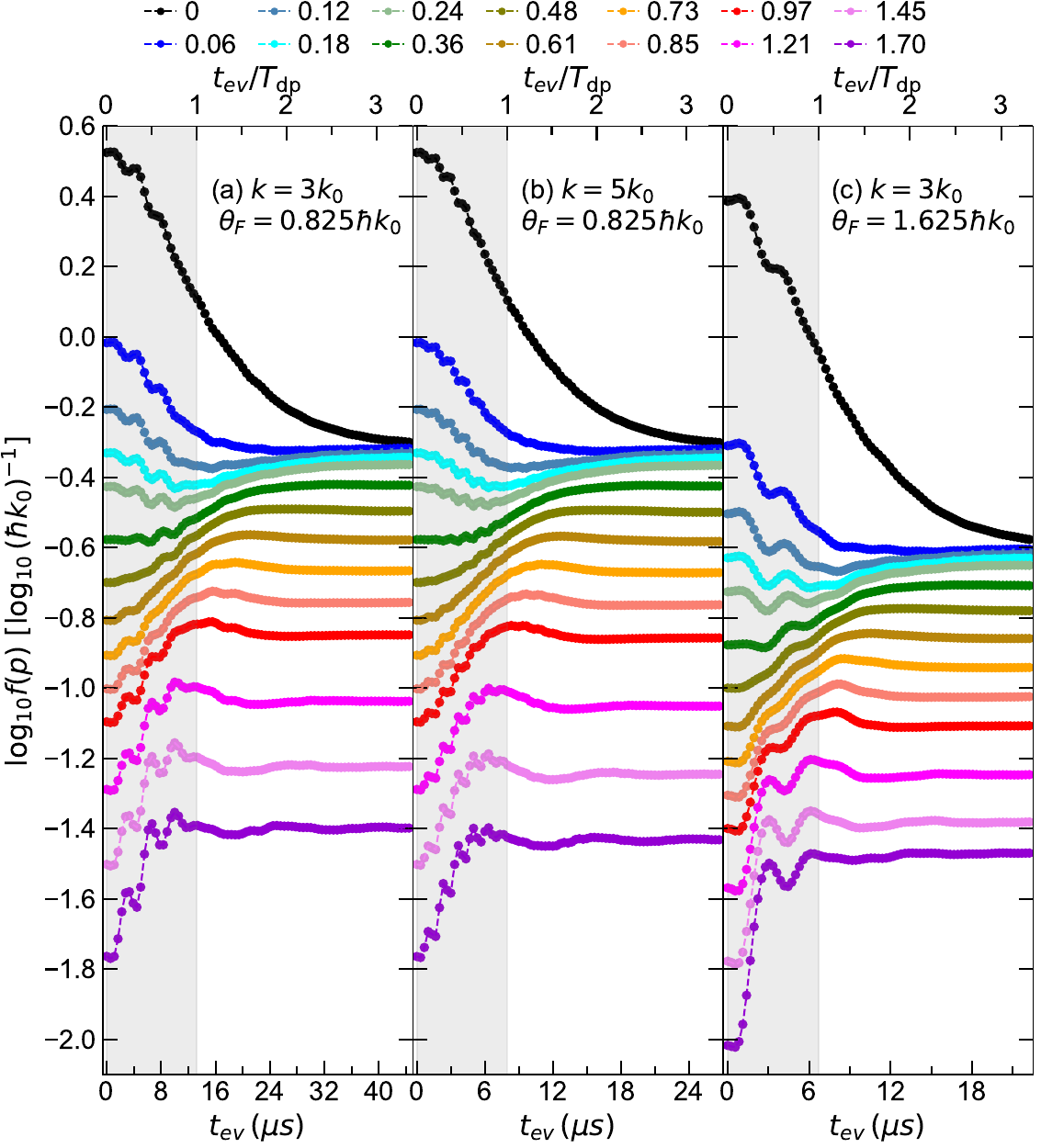}
    \vspace{-0.15cm}
    \caption{{\it Time evolution of $f(p)$ in and around the central peak for three-peak states.} The initial states have: (a) $N=33$ ($\theta_F=0.825\hbar k_0$) and $k=3 k_0$, (b) $N=33$ ($\theta_F=0.825\hbar k_0$) and $k=5 k_0$, and (c) $N=65$ ($\theta_F=1.625\hbar k_0$) and $k=3 k_0$. Note the change in the times shown in each panel. The upper $x$ axis shows the dimensionless quantity $t_{\rm ev}/T_{\rm dp}$. The legends show the value of $p/\theta_{\rm F}$. The vertical gray bands ($t_{\rm ev}\in[0,1]T_{\rm dp}$) highlight the timescale in which the hydrodynamization oscillations damp out.}
    \label{fig:3peakfpvsp}
\end{figure} 

How the occupation $f(p)$ of each momentum mode approaches the GGE prediction depends in a complicated way on $p$, $k$, and $\theta_F$. Figure~\ref{fig:3peakfpvsp} contains most of the features we have discussed with regard to the three timescales. It shows the time evolution of $f(p)$ for $p<2\theta_F$ (i.e., within the central momentum peak), for $N=33$ when $k=3 k_0$ [Fig.~\ref{fig:3peakfpvsp}(a)] and when $k=5 k_0$ [Fig.~\ref{fig:3peakfpvsp}(b)], and for $N=65$ when $k=3 k_0$ [Fig.~\ref{fig:3peakfpvsp}(c)]. Note that the absolute time axes differ across the panels so that the dimensionless (upper) time axes $t_{\rm ev}/T_{\rm dp}$ [see Eq.~\eqref{eq:tdp}] are the same. We will now enumerate the take home messages from this figure:
\begin{itemize}
    \item[1.] The period of the hydrodynamization oscillations does not scale as $T_{\rm dp}$. It is $T^c_{\rm hd}$ for all curves, as per Eq.~\eqref{eq:thd}. This is also seen in Fig.~\ref{fig:3peakek}, where the dimensionless time is more apt for this comparison.
    \item[2.] The damping of the hydrodynamization oscillations in all panels and for all $p$ occurs within $t_{\rm ev}\simeq T_{\rm dp}$ (highlighted by the shaded bands), as per Eq.~\eqref{eq:tdp}. Note that the detailed ways in which the oscillations damp depend on $f(p)$, but the characteristic times for each $f(p)$ scales with $T_{\rm dp}$.
    \item[3.] The local prethermalization times, i.e., how long it takes for $f(p)$ to equilibrate, also scale with $T_{\rm dp}$, as can be seen by comparing curves of the same color.
    \item[4.] The diversity of shapes among the $f(p)$ curves in Fig.~\ref{fig:3peakfpvsp} makes it difficult to extract a prethermalization time for different values of $p$. However, it is clear that $T_{\rm lp}(f(p))$ steadily decreases as $p$ increases. The continuous change in these times reflects an important characteristic of local prethermalization, that it leads to redistribution of occupancies only among nearby momentum groups. This contrasts with hydrodynamization oscillations, where the amplitudes get redistributed among distant momentum groups. 
    \item[5.] The longest prethermalization times (most clearly seen in the black curves) are longer than the longest hydrodynamization damping times. While for some observables these timescales might be the same, $T^d_{\rm hd}(\hat O)$ is generally a lower bound for $T_{\rm lp}(\hat O)$. In the experiments in Ref.~\cite{le2023observation}, $T_{\rm lp}(f(p))$ was about three times longer than the theoretical $T_{\rm lp}(f(p))$ for $p$ within the central peak, while $T^d_{\rm hd}(f(p))$ was about the same.
    \item[6.] For $p\gtrsim\theta_F$, $T_{\rm lp}(f(p))\simeq T^d_{\rm hd}(f(p))\simeq T_{\rm dp}$. The saturation as $p$ increases is understandable as neither local prethermalization nor hydrodynamization damping can proceed on a timescale that is faster than $T_{\rm dp}$.
\end{itemize}

\begin{figure}[!t]
    \includegraphics[width=0.985\columnwidth]{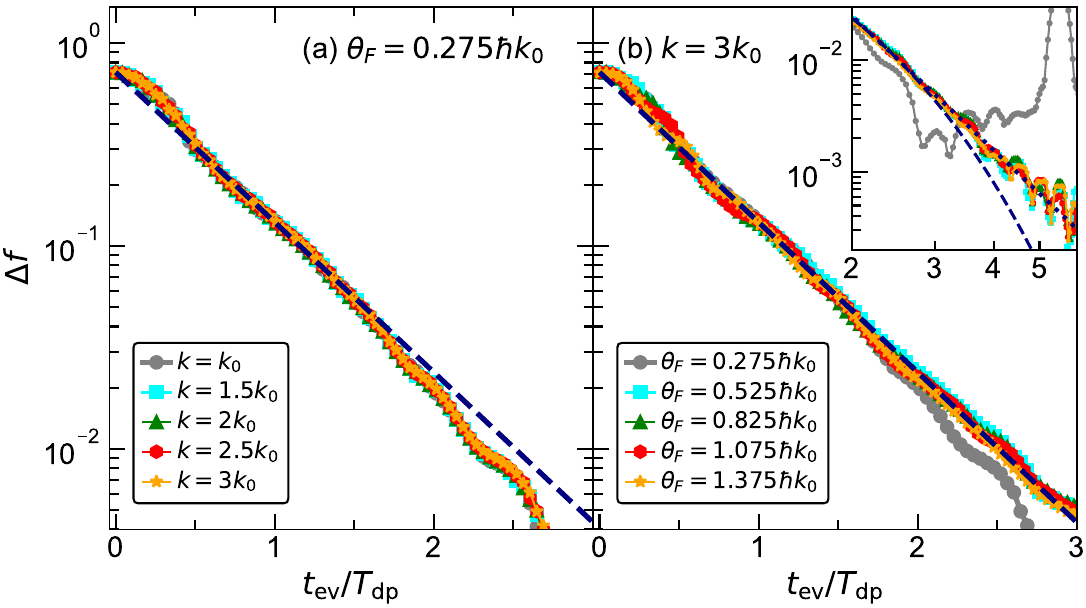}
    \vspace{-0.15cm}
    \caption{{\it Prethermalization of $f(p)$ for three-peak states.} $\Delta f$ vs the dimensionless time $t_{\rm ev}/T_{\rm dp}$. (a) We fix $N=11$ ($\theta_F=0.275\hbar k_0$) and show results for different $k$. (b) We fix $k=3k_0$ and show results for different $N$ ($\theta_F$). The dashed lines show the result of a fit $a\exp(-b\,t_{\rm ev}/T_{\rm dp})$ to the results for the largest number of particles in (b), which yields $b=1.7$. Inset of (b): same as (b) but plotted on a log-log scale for later times. The dashed line is the same exponential fit as in the main panels. The dotted line is a power law guide to the eye, $(t_{\rm ev}/T_{\rm dp})^{-4}$.  The $N=11$ ($\theta_F=0.275\hbar k_0$) curve exhibits a revival after $t_{\rm ev}=5 T_{\rm dp}$, a finite size effect resulting from the small number of particles.}
    \label{fig:3peakDfvst}
\end{figure}

Building on the observations from Fig.~\ref{fig:3peakfpvsp}, we next show that the prethermalization timescale for the entire momentum distribution is also proportional to $T_{\rm dp}$. To characterize the local prethermalization of the momentum distribution, we calculate the relative difference
\begin{equation}\label{eq:diffDdef}
    \Delta f(t_{\rm ev})= \int dp\, |f(p,t_{\rm ev})-\bar f(p)|,
\end{equation}
where $\bar f(p)$ is the locally equilibrated momentum
distribution (see Appendix~\ref{app:initialstate}).

In Fig.~\ref{fig:3peakDfvst}, we plot $\Delta f(t_{\rm ev})$ vs $t_{\rm ev}/T_{\rm dp}$ for systems with the same number of particles ($N=11$) and different $k$ [Fig.~\ref{fig:3peakDfvst}(a)], and for systems with the same value of $k=3k_0$ and different $N$ [Fig.~\ref{fig:3peakDfvst}(b)]. In both panels, one can see that the curves of $\Delta f$ vs $t_{\rm ev}/T_{\rm dp}$ collapse onto each other, except for the $N=11$ curve past $t_{\rm ev}\sim 1.5 T_{\rm dp}$ in Fig.~\ref{fig:3peakDfvst}(b), which deviates due to finite-size effects. For times $0.5 T_{\rm dp} \lesssim t_{\rm ev}\lesssim 3 T_{\rm dp}$, the results for $N>11$ are well described by an exponential function. Specifically, $\Delta f(t)\propto \exp(-b\,t_{\rm ev}/T_{\rm dp})$, which means that one can define a prethermalization time $T_{\rm lp}$ for the entire momentum distribution that is proportional to $T_{\rm dp}$. At long times [see the inset in Fig.~\ref{fig:3peakDfvst}(b)] $\Delta f(t)$ decays polynomially: $\Delta f(t) \propto (t_{\rm ev}/T_{\rm dp})^{-\alpha}$ with $\alpha\approx 4$.

It has previously been argued that the quantum dynamics of TG gas momentum distributions can be like those of generic observables in interacting integrable models that cannot be mapped onto noninteracting models~\cite{zhang_vidmar_22}. We therefore expect our TG results to qualitatively describe dynamics away from the TG limit, where the Lieb-Liniger model cannot be mapped onto a noninteracting model, with a proportionality constant between $T_{\rm lp}$ and $T_{\rm dp}$ that depends on the contact interaction strength $g$ [see Eq.~\eqref{eq:H_lieb_liniger}]. For completeness, in Appendix~\ref{app:nopulse_3p} we study the dynamics of the density distribution and connect the results for local prethermalization obtained in this section to those in Ref.~\cite{le2023observation}.

\section{Two-peak states}\label{sec:2peak}

The rapidity distribution in Eq.~\eqref{eq:3peaks} is multimodal, with each quasiparticle in a superposition of plane waves with rapidities $0$ and $\pm 2\hbar k$. For such a rapidity distribution, the quasiparticle energy distribution is bimodal, with two peaks centered at energies $0$ and $(2\hbar k)^2/2m$. As discussed in Sec.~\ref{sec:hydro}, and as shown in Sec.~\ref{sec:3peak}, the energy difference between the center of the latter two peaks is what determines $T^c_{\rm hd}$. Therefore, we expect that if the quasiparticles after the quench are in superpositions of plane waves with similar energies, say with rapidities $\pm 2\hbar k$ [corresponding to energies centered at $(2\hbar k)^2/2m$], there will be no hydrodynamization. In that case, the state after the quench is still ``high-energy'' compared to the ground state, since $(2\hbar k)^2/2m$ remains the highest energy scale, but there are no dynamics associated with that energy scale. The goal of this section is to confirm this expectation.

We study the dynamics after removing the central ($k=0$) peak in Eq.~\eqref{eq:3peaks}, so that we have bimodal rapidity distributions with $A_{\pm 1}=1/\sqrt{2}$: 
\begin{equation}\label{eq:2peakwf}
    |\Psi_{\rm 2p}(t_{\rm ev}=0)\rangle=\prod_{|\theta|<\theta_F}\frac{1}{\sqrt{2}}(\hat c^{\dagger}_{\theta+2\hbar k}+\hat c^{\dagger}_{\theta-2\hbar k})|0\rangle\,. 
\end{equation}
There is no hydrodynamization timescale associated with these ``two-peak'' states. Dephasing of rapidities still occurs, leading to the evolution of observables. Dephasing results from the energy differences between the quasiparticles with rapidities $\theta-2\hbar k$ and $\theta+ 2\hbar k$, which have different values within a bandwidth $W'$, which is twice as large as in the three-peak case $W'=2W=16\hbar k \theta_F/(2m)$. The dephasing timescale is thus twice as fast for any given $k$ and $\theta_F$ 
\begin{equation}\label{eq:tdpp}
    T'_{\rm dp}=\frac{2\pi\hbar}{W'}=\frac{\pi m}{4 k\theta_F}.
\end{equation}

\begin{figure}[!t]
    \includegraphics[width=0.985\columnwidth]{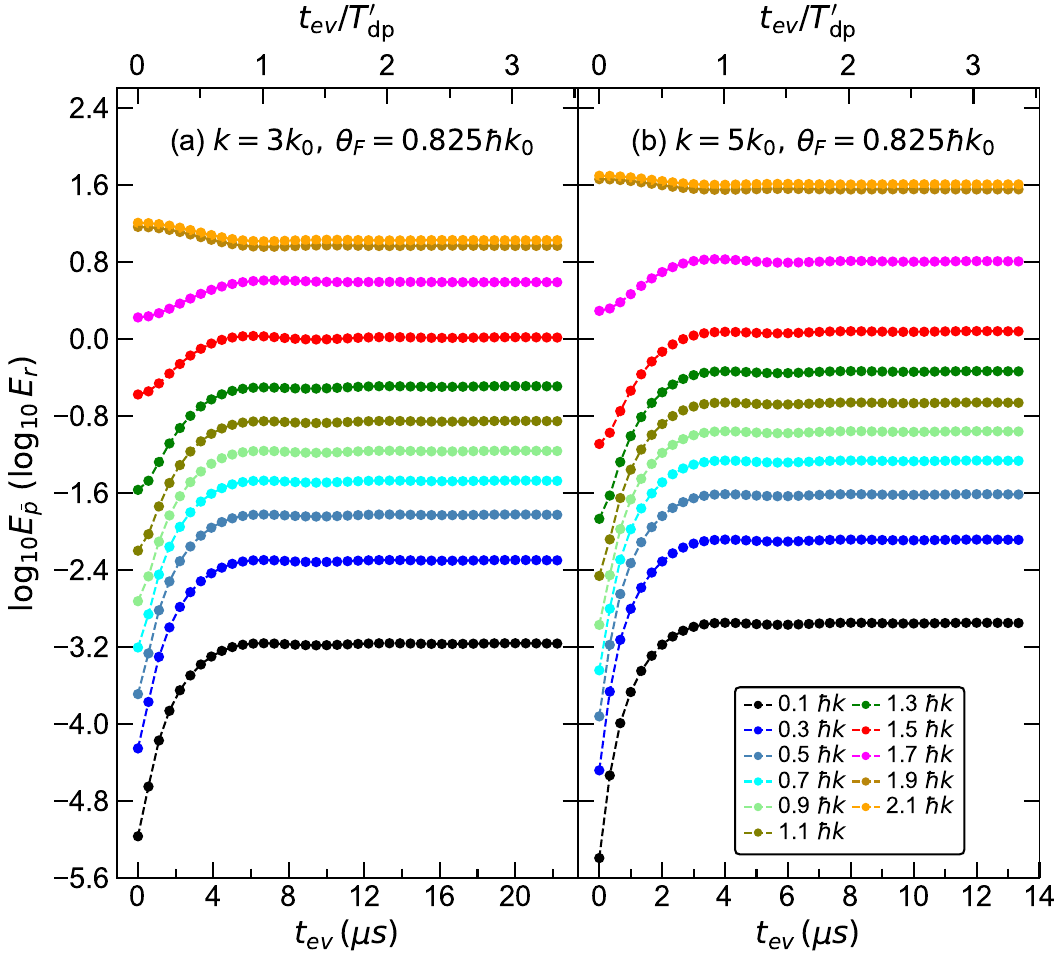}
    \vspace{-0.15cm}
    \caption{{\it Time evolution of $E_{\bar p}$ for the ``two-peak state" [see Eq.~\eqref{eq:2peakwf}].} We show results for a system with $N=33$ atoms ($\theta_F=0.825$) for (a) $k=3k_0$ and (b) $k=5k_0$. Each curve is integrated within a momentum interval of 0.2 $\hbar k$. Note the change in the times shown in both panels. The upper $x$ axis shows the dimensionless quantity $t_{\rm ev}/T_{\rm dp}'$.}
    \label{fig:2peakek}
\end{figure}

In Fig.~\ref{fig:2peakek}, we plot the time evolution of $E_{\bar p}$ for two-peak states for two values of $k$, as we did in Fig.~\ref{fig:3peakek} for three-peak states. There are, as expected, no hydrodynamization oscillations. All one can see in Fig.~\ref{fig:2peakek} is local prethermalization; comparing the same momentum curves in the two panels we see that it occurs in a time that is proportional to $T'_{\rm dp}$ (see the top $x$ axis scale). In fact, the local prethermalization time is approximately the same for all momentum groups, as it is for high-momentum modes in the three-peak case. These results highlight the necessity of multimodal energy distributions for hydrodynamization to occur. They also highlight that local prethermalization can occur in remarkably short times even in the absence of hydrodynamization, because the local prethermalization time is inversely proportional to $k$, which is not related to any microscopic scale associated with the Hamiltonian of the many-body system.

\begin{figure}[!t]
\includegraphics[width=0.985\columnwidth]{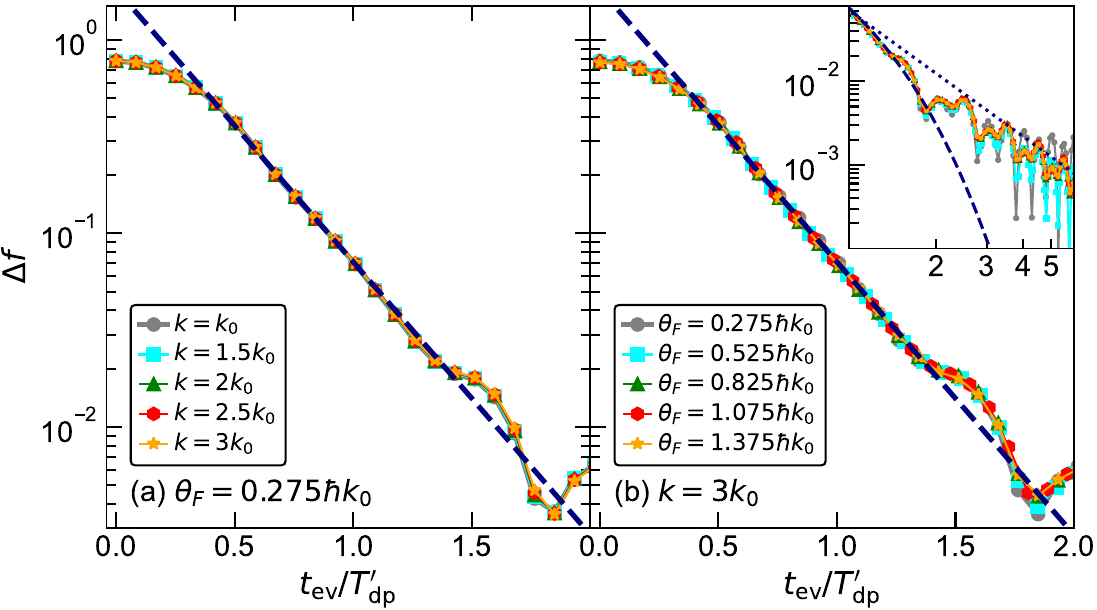}
    \vspace{-0.15cm}
    \caption{{\it Local prethermalization of $f(p)$ for the ``two-peak'' state.} $\Delta f$ vs the dimensionless time $t_{\rm ev}/T'_{\rm dp}$. (a) We fix $N=11$ ($\theta_F=0.275\hbar k_0$) and show results for different $k$, as in Fig.~\ref{fig:3peakDfvst}(a). (b) We fix $k=3k_0$ and show results for different $N$ ($\theta_F$), as in Fig.~\ref{fig:3peakDfvst}(b). The dashed lines show a fit $a\exp(-b\,t_{\rm ev}/T'_{\rm dp})$ to the results with largest atom number in (b) and $t_{\rm ev}\in(0.5,1.5)T'_{\rm dp}$, which yields $b=3.2$. Inset of (b): same as (b) but plotted on a log-log scale for later times. The dashed line is the same exponential fit as in (b). The dashed line is a guide to the eye and shows power-law behavior, $(t_{\rm ev}/T'_{\rm dp})^{-2.5}$.}
    \label{fig:2peakDfvst}
\end{figure}

In Fig.~\ref{fig:2peakDfvst} we report our results for $\Delta f$ vs $t_{\rm ev}/T'_{\rm dp}$ for two-peak states (see Appendix~\ref{sec:nopulse_2p} for the corresponding results for the dynamics of the density distribution). They parallel the results for the three-peak states shown in Fig.~\ref{fig:3peakDfvst}. The main difference is that (as expected) the prefactor $b$ of the dimensionless time in the exponential decay of $\Delta f$ in Fig.~\ref{fig:2peakDfvst} is about twice that in Fig.~\ref{fig:3peakDfvst}, while the dimensionless time during which this exponential decay is visible in Fig.~\ref{fig:2peakDfvst} is about half that in Fig.~\ref{fig:3peakDfvst}. In both figures, a slower power-law decay emerges at late times when $\Delta f$ is smaller than a fraction of a percent. Given the smallness of $\Delta f$ at those times, such a slower decay is probably irrelevant to experimental observations, but it might be of interest to explore theoretically in the future.

\subsection{Symmetric two-peak states}\label{sec:sym2peak}

\begin{figure}[!t]
\includegraphics[width=0.985\columnwidth]{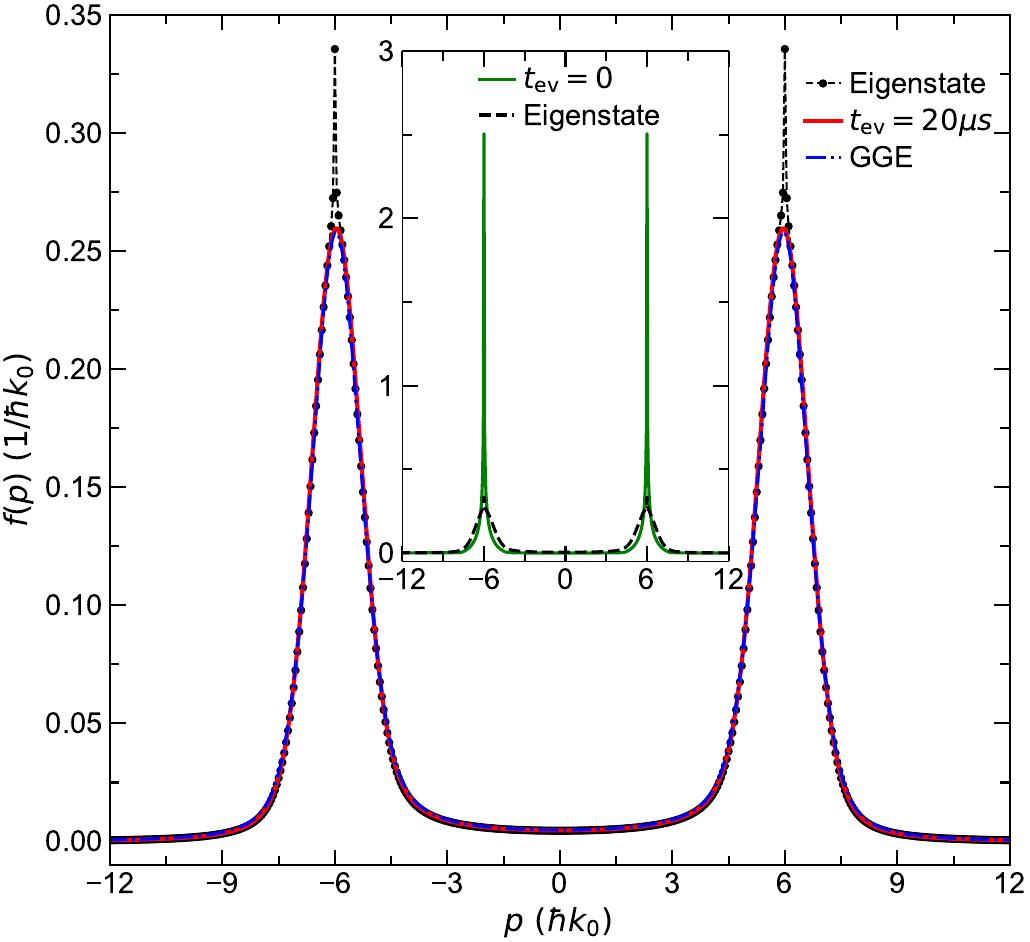}
    \vspace{-0.15cm}
    \caption{{\it Momentum distributions for two-peak states.} Main panel: $f(p)$ for the two-peak postquench wave function [Eq.~(\ref{eq:2peakwf}), with $k=3k_0$ and $N=33$] at $t_{\rm ev}=20$ $\mu$s (red solid line), the corresponding symmetric two-peak state [Eq.~(\ref{eq:2peakeigenstate})] (black dots; the dashed line is a guide to the eye), and the GGE prediction (blue dashed-dotted line). Inset: $f(p)$ for the two-peak postquench wave function [Eq.~(\ref{eq:2peakwf}), with $k=3k_0$ and $N=33$] at $t_{\rm ev}=0 $ $\mu$s (green solid line) and the corresponding symmetric two-peak state [Eq.~(\ref{eq:2peakeigenstate})] (black dashed line). The small black dots in the main panel mark the quantized momenta in our finite system.}
    \label{fig:2peakeigenstate}
\end{figure}

The contrast between our three- and two-peak results makes it clear that the time evolution of observables depends crucially on the rapidity distributions. For two-peak states, the relative phases of the eigenstates also matter greatly. To illustrate this point, we constructed the following many-quasiparticle states
\begin{equation}\label{eq:2peakeigenstate}
    |\Psi_{\rm e}\rangle=\prod_{|\theta|<\theta_F}\frac{1}{\sqrt{2}}(\hat c^{\dagger}_{\theta+2\hbar k}+\hat c^{\dagger}_{-\theta-2\hbar k})|0\rangle\,.
\end{equation}
They have the same rapidity distribution as the states in Eq.~\eqref{eq:2peakwf}, yet because each term in the product state consists of a superposition of equal energy states, such symmetric two-peak states exhibit no dynamics; they are eigenstates of the translationally invariant TG gas. Also, despite having the same rapidity distributions, the states in Eqs.~\eqref{eq:2peakwf} and~\eqref{eq:2peakeigenstate} have different momentum distributions, as shown in the inset in Fig.~\ref{fig:2peakeigenstate}. 

Because of generalized eigenstate thermalization~\cite{cassidy_clark_11, caux_essler_13, Pozsgay_2014, vidmar2016generalized, mori_19, dymarski_19}, the momentum distribution of the state in Eq.~\eqref{eq:2peakwf} is expected to equilibrate to the same momentum distribution as all energy eigenstates that have the same rapidity distribution, to within differences that vanish in the thermodynamic limit. Our results in Fig.~\ref{fig:2peakeigenstate} confirm this. The asymptotic $f(p)$ (at $t_{\rm ev}=20$ $\mu$s) from the superposition of Eq.~\eqref{eq:2peakwf} exactly coincides with the GGE prediction. The unchanging momentum distribution from the symmetric two-peak state [Eq.~\eqref{eq:2peakeigenstate}] is also indistinguishable from the GGE prediction at all momenta, except at the singular $|p|=2\hbar k$ points and the nearest momentum points to it. As in many other systems studied in the literature (see Ref.~\cite{vidmar2016generalized} for a review), we expect the normalized differences between the equilibrated, the GGE, and the symmetric two-peak state momentum distributions, which are small but finite in our finite systems, to vanish in the thermodynamic limit. 

\section{Quantum Newton's cradle}\label{sec:qnc}

States with two dominantly populated Bragg peaks, like the ones considered in the previous section, can be realized experimentally by applying a quantum Newton's cradle (QNC) sequence~\cite{wu2005splitting, kinoshita2006quantum}. The QNC sequence contains two Bragg pulses like the ones in Eq.~(\ref{eq:U_pulse}). Optimal QNC pulses are $\pi/(4\sqrt{2}\omega_r)$ long (23 $\mu$s for $k_0$), separated by $\pi/(4\omega_r)$ (33 $\mu$s for $k_0$), where $\omega_r=E_r/\hbar$, $E_r=\hbar^2k_0^2/(2m)$ is the recoil energy~\cite{wu2005splitting, kinoshita2006quantum}. These pulses are a large enough fraction of $T^c_{\rm hd}$ that significant hydrodynamization can occur during them. To reduce this effect, we use asymmetric QNC pulses. A long pulse followed by a short pulse still largely depletes the central peak, but at the cost of requiring a higher pulse amplitude, which compromises the two-level approximation that underlies the theory~\cite{wu2005splitting} and leads to a larger excitation of higher Bragg peaks. The advantage, however, is that there is less many-body evolution during the second pulse, so that the evolution after the second pulse presents a cleaner test of the absence of hydrodynamization. For simplicity, we set the amplitude of both pulses to be the same $U_{\rm pulse}=18$ $E_r$, and label the duration of the first (second) Bragg pulse as $t_{\rm pulse;1}$ ($t_{\rm pulse;2}$). We calculate these times assuming an ideal two-level system~\cite{wu2005splitting}, which ignores many-body interactions. The ideal pulse separation ($t_{\rm mid}$) is the same as it is for symmetric pulses.

\begin{figure}[!t]
    \includegraphics[width=0.985\columnwidth]{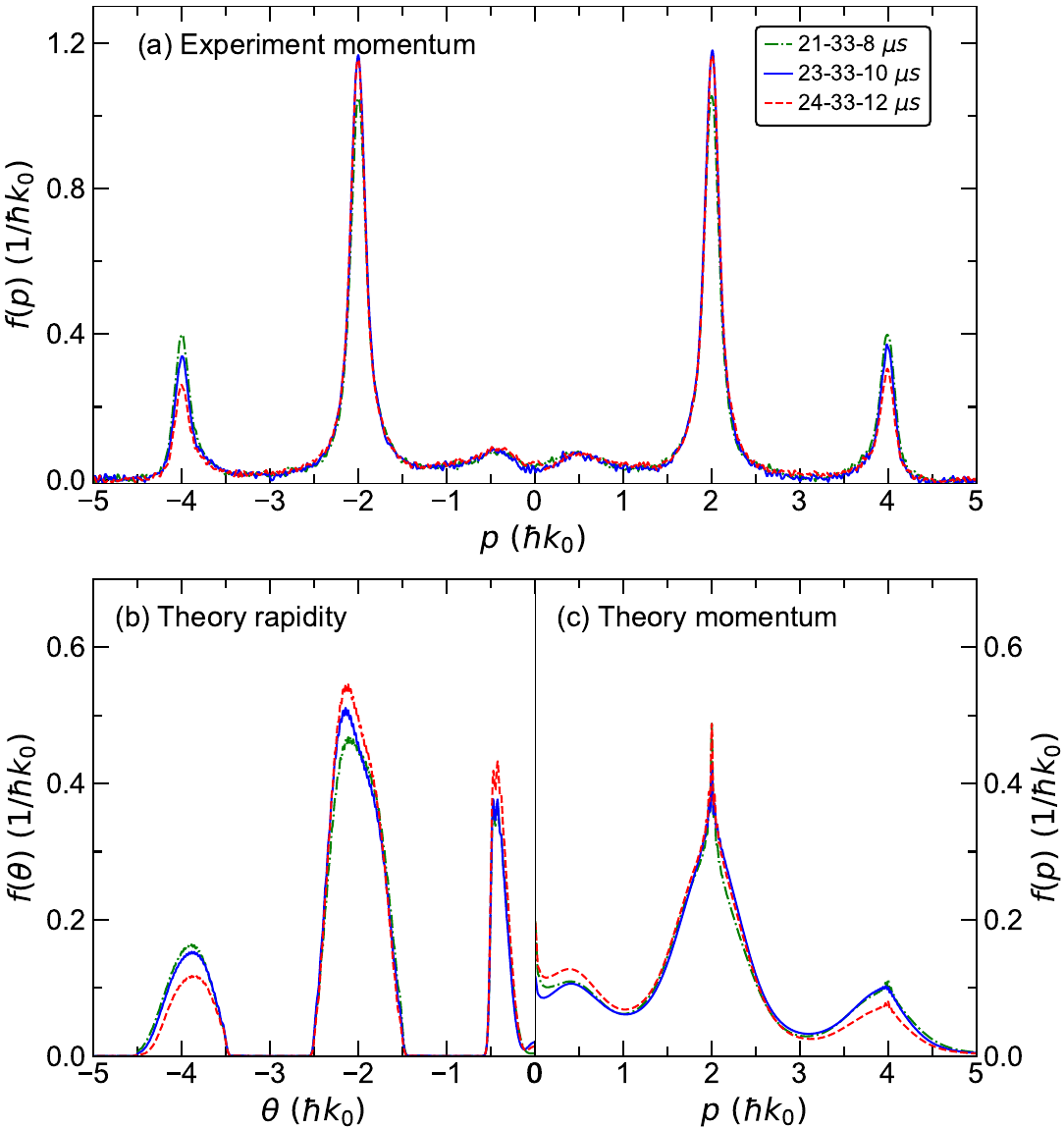}
    \vspace{-0.15cm}
    \caption{{\it Momentum and rapidity distributions at the end of different QNC sequences.} Results are shown after three sequences with: $t_{\rm pulse;1}=21$ $\mu$s, $t_{\rm mid}=33$ $\mu$s, and $t_{\rm pulse;2}=8$ $\mu$s (green dashed-dotted line); $t_{\rm pulse;1}=23$ $\mu$s, $t_{\rm mid}=33$ $\mu$s, and $t_{\rm pulse;2}=10$ $\mu$s (blue solid line); and $t_{\rm pulse;1}=24$ $\mu$s, $t_{\rm mid}=33$ $\mu$s, and $t_{\rm pulse;2}=12$ $\mu$s (red dashed line). (a) Experimental results for the momentum distribution. (b), (c) Corresponding theoretical results for the rapidity (b) and momentum (c) distribution of a single 1D gas in the TG limit.}
    \label{fig:QNCfp}
\end{figure}

The length of the second pulse represents a trade-off between it being short and minimizing the $|n|=2$ excitations. Because many-body interactions complicate the ideal two-level system results in a way that is hard to calculate, we empirically determine a suitable QNC sequence. We start from the same pre-quench state (which is also the same as for the single Bragg pulse discussed Appendix~\ref{app:nsbp}), and change the duration (and amplitude) of the two Bragg pulses. Figures.~\ref{fig:QNCfp}(a)--\ref{fig:QNCfp}(c) show respectively results for the momentum distributions obtained experimentally, and the rapidity and momentum distributions obtained theoretically, each after three different QNC sequences. The theoretical momentum distributions are qualitatively similar to the experimental ones, although the theory exhibits lower and broader peaks than the experiment. Of the three QNC sequences, we choose to work with $t_{\rm pulse;1}=23$ $\mu$s, $t_{\rm mid}=33$ $\mu$s, and $t_{\rm pulse;2}=10$ $\mu$s (blue solid line). When $t_{\rm pulse;2}$ is shorter, the occupation of the $n=\pm 1$ Bragg peaks starts to drop in favor of higher $|n|$ peaks. Still, one can see that both in the experiment and the theory some atoms [and quasiparticles, as shown in Fig.~\ref{fig:QNCfp}(b)] populate the $n=\pm 2$ Bragg peaks for our chosen sequence. We further note from Fig.~\ref{fig:QNCfp}(b) that although the QNC sequences transfer nearly all quasiparticles with $\theta\sim 0$ to higher rapidity modes, some quasiparticles with rapidities $\theta \lesssim \theta_F$ remain. They are the reason there are small low-momentum side peaks in the experimental and theoretical momentum distributions about $p=0$.

\begin{figure}[!t]
    \includegraphics[width=0.985\columnwidth]{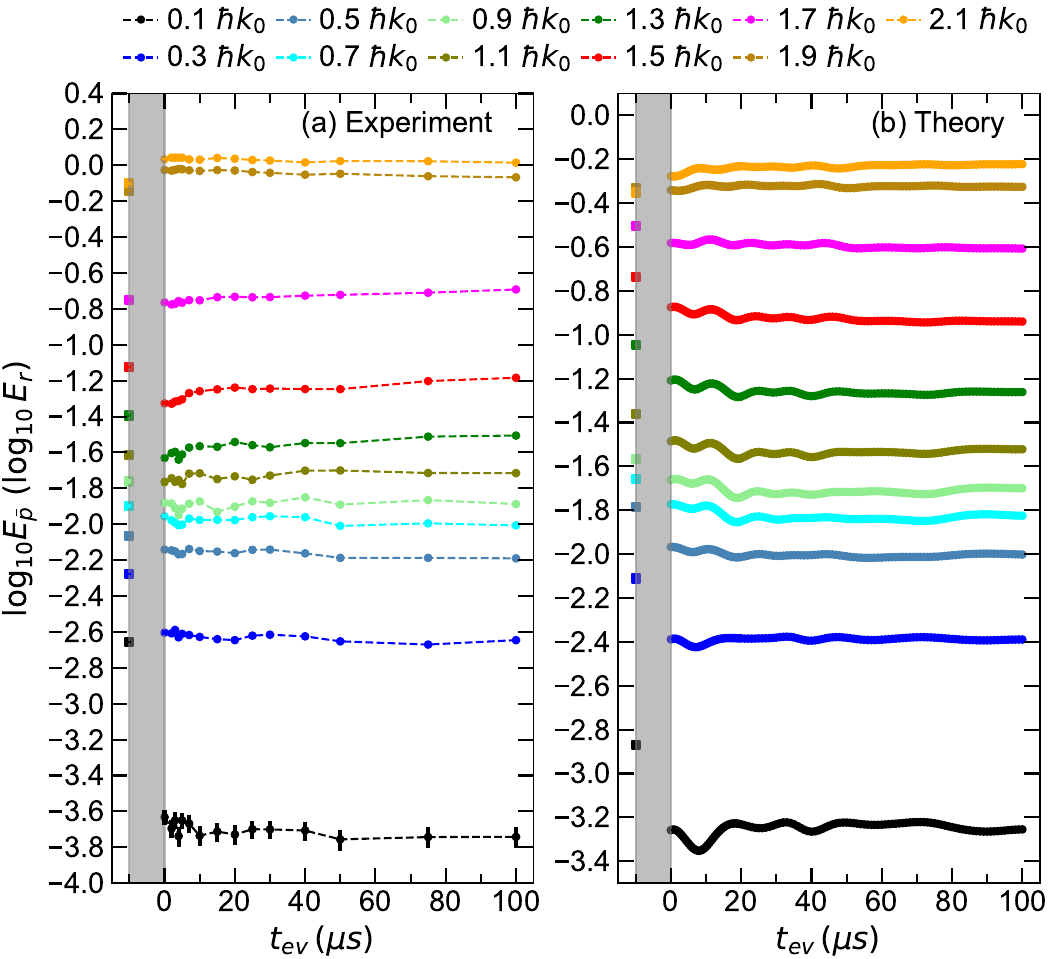}
    \vspace{-0.15cm}
    \caption{{\it Suppression of hydrodynamization after a QNC sequence.} The curves show the time evolution of $E_{\bar p}$, in $0.2\hbar k_0$ wide momentum intervals, following the QNC sequence with $t_{\rm pulse;1}=23$ $\mu$s, $t_{\rm mid}=33$ $\mu$s, and $t_{\rm pulse;2}=10$ $\mu$s. On the $x$ axis, $t_{\rm ev}=0$ is the time at which the second Bragg pulse ends. We also plot, at $t_{\rm ev}=-10$ $\mu$s, the values of $E_{\bar p}$ at the beginning of the second pulse (the second pulse duration is indicated by the vertical gray band). (a) Experimental results for $\bar \gamma_0=3.4$ (see Appendix~\ref{app:exp-the-bra} for the definition of $\bar \gamma_0$). (b) Corresponding theoretical results for a single 1D gas in the TG limit, with the same trap parameters and energy density per particle as the experimental array of 1D gases. Dashed lines are guides to the eye.}
    \label{fig:expQNC}
\end{figure}

In Fig.~\ref{fig:expQNC}, we plot the time evolution of $E_{\bar p}$ after the 23-33-10 $\mu$s QNC sequence in the experiment [Fig.~\ref{fig:expQNC}(a)] and the corresponding theoretical results for a single 1D gas in the TG limit [Fig.~\ref{fig:expQNC}(b)]. The gray shaded area indicates the time of the second ($t_{\rm pulse;2}=10$ $\mu$s) Bragg pulse in the sequence, which has the same amplitude and duration as the single Bragg pulse after which the dynamics are shown in Fig.~\ref{fig:exp10mus} in Appendix~\ref{app:nsbp}. Figure~\ref{fig:expQNC} also shows, at $t_{\rm ev}=-10$ $\mu$s, the values of $E_{\bar p}$ at the beginning of the second pulse. The dominant effect of the second pulse is to deplete the $n=0$ peaks in the rapidity and momentum distributions. It also affects the occupation of the momentum modes between the $n=0$ and $|n|=1$ peaks, partially undoing the hydrodynamization evolution after the first pulse. Qualitatively, the system is far from equilibrium after the second pulse, as it would be after a single pulse of the same length. It just lacks the $n=0$ peak.

\begin{figure}[!t]
    \includegraphics[width=0.985\columnwidth]{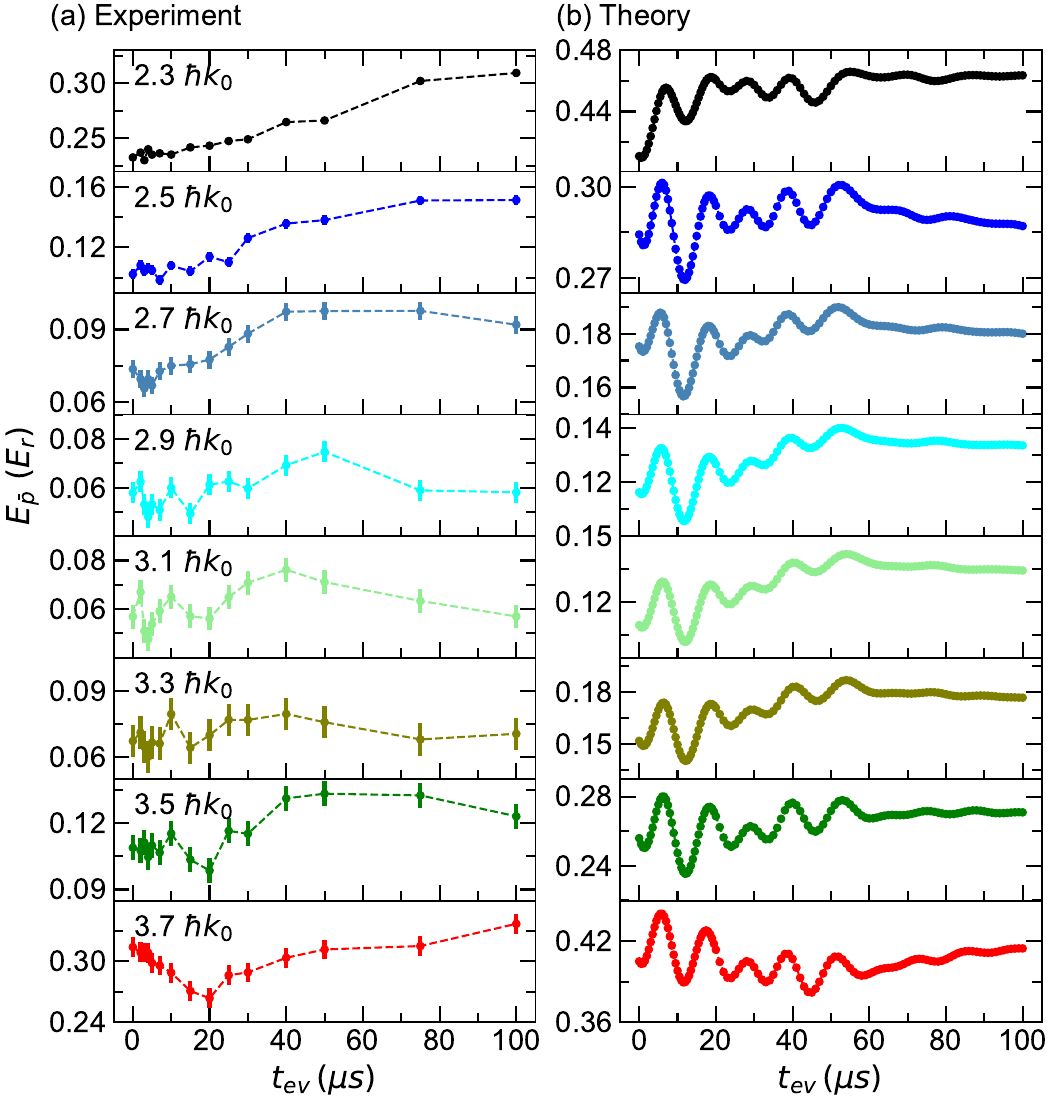}
    \vspace{-0.15cm}
    \caption{{\it Hydrodynamization at higher momenta.} Time evolution of $E_{\bar p}$ [see Eq.~\eqref{eq:barEp}] between $2\hbar k_0$ and $4\hbar k_0$. The QNC sequence is the same as in Fig.~\ref{fig:expQNC}, and we use $0.2\hbar k_0$ wide momentum intervals to compute $E_{\bar p}$. (a) Experimental results for $\bar \gamma_0=3.4$. (b) Corresponding theoretical results for a single 1D gas in the TG limit, with the same trap parameters and energy density per particle as the experimental array of 1D gases. Dashed lines are guides to the eye.}
    \label{fig:expQNClargep}
\end{figure}

The experimental and the theoretical results for $E_{\bar p}$ in Figs.~\ref{fig:expQNC}(a) and~\ref{fig:expQNC}(b), respectively, are qualitatively similar to each other for $t_{\rm ev}>0$. The curves are all relatively flat, with no oscillations with $T^c_{\rm hd}/2=33$ $\mu$s, in stark contrast to the results due to the single pulse in Fig.~\ref{fig:exp10mus}. We therefore confirm experimentally that when the central peak is suppressed there is negligible hydrodynamization.

The curves in Fig.~\ref{fig:expQNC} feature small oscillations with a higher frequency than that in Fig.~\ref{fig:exp10mus}. Such short timescale momentum mode changes are caused by the presence of atoms with $n=\pm 2$ (see Fig.~\ref{fig:QNCfp}). In Fig.~\ref{fig:expQNClargep}, we show our experimental [Fig.~\ref{fig:expQNClargep}(a)] and theoretical [Fig.~\ref{fig:expQNClargep}(b)] results for $E_{\bar p}$ of momentum modes between $2\hbar k_0$ and $4\hbar k_0$. At those energies, we find a stronger hydrodynamization signal at the frequency associated with the difference between the energies of the $n=\pm 2$ and $\pm 1$ Bragg peaks, which is three times higher than that between the $n=\pm 1$ and $0$ Bragg peaks. The theoretical data in particular, which have better time resolution, exhibit clear oscillations of $E_{\bar p}$ with the expected frequency.

\section{Heavy-ion collisions} \label{sec:heavyion}

We next apply the lessons we have learned about hydrodynamization and local prethermalization to relativistic heavy-ion collisions. To avoid confusion in the application of these ideas to particle physics, it is important to make clear that the term ``rapidities'' used in this paper and in the integrable systems literature (denoting the momenta of quasiparticles) has no relation to rapidities in high energy physics (denoting transverse relativistic momenta). The two and three peak quenches of 1D gases that we consider in this paper resemble collisions between atomic groups with distinct momenta, with the provisos that each single atom's wave function is part of all colliding groups and there is significant entanglement among individual atoms. The fact that the ``colliding'' parts of the wave function spatially overlap from the very beginning of the quench is critical to their value in understanding short-time evolution. The short-time behavior would be difficult to extract in a conventional collision between incident clouds of atoms, as it would happen throughout the process of the clouds spatially merging. 

It will become apparent below that the time it takes for colliding relativistic nuclei to overlap is generally shorter than the time it takes for hydrodynamization to run its course. To estimate the time it takes for the nuclei to overlap, note that the highest collision energies probed at the Relativistic Heavy Ion Collider (RHIC) and the Large Hadron Collider (LHC) have $\gamma_L=1/\sqrt{1-v^2/c^2}$ Lorentz factors of about 100 and 2500, respectively. In the laboratory frame, the nuclei look like space contracted disks of $D/\gamma_L$, where $D$ is the nuclear diameter ($D\sim 14$ fm for heavy nuclei), traveling at about the speed of light, $c$. The two nuclei are fully overlapped in a head-on collision within $\simeq 0.14$ fm/$c$ at RHIC and $\simeq 0.0056$ fm/$c$ at the LHC. We take the point of full overlap as $t=0$ for considering postquench evolution, analogous to the time at the end of a Bragg-pulse quench of 1D gases.

The colliding nuclei form a quark-gluon plasma, which grows as the nuclei recede from each other. The system is far from integrable, so its large-distance and long-time dynamics are described by hydrodynamics with a very small ratio between the shear viscosity $\eta$ and the entropy density $s$, $\eta/s\approx 1/(4\pi)$. ($\eta/s = 1/(4\pi)$ is the ratio found in infinitely strongly coupled quantum theories described by a holographic dual gravitation theory~\cite{policastro_01}.) This small ratio implies that the quark-gluon plasma is a strongly coupled fluid, unlike the asymptotically free quarks and gluons at much higher energies~\cite{busza_18}. That is, the quark-gluon plasma is an interacting many-body quantum system, amenable to the approach to studying dynamics we use in this paper. A major open question in the description of heavy-ion collisions is why hydrodynamics works at very short times after the collision. The proper times $\tau^{\rm exp}_{0}$ at which hydrodynamic descriptions start to work depend on the specific model used in calculations; they are around $0.4–1.0$ fm/$c$ for the RHIC and $0.2–0.6$ fm/$c$ for the LHC~\cite{busza_18}. Our goal here is to provide insight into this open question.

We assume that there is a quasiparticle description of the quark-gluon plasma. Although the quark-gluon plasma is not integrable, these quasiparticles do not decay on hydrodynamization timescales, which are the shortest timescales. We label the quasiparticle masses $m_{\rm qs}$. Although these masses are unknown, we assume that they are smaller than the {\it constituent quark} masses of the $u$ and $d$ quarks ($\sim$300 MeV/$c^2$) and larger than the {\it current quark} masses ($\sim$3 MeV/$c^2$). The two orders of magnitude difference between {\it current quark} and {\it constituent quark} masses are due to the gluon field~\cite{cabo_rigol_02, rigol_cabo_00}. The rapidity distribution is also unknown, but if there has not been enough time for significant many-body evolution, then the longitudinal rapidity distribution will contain peaks at $\pm \theta_{\gamma_L}=\pm \gamma_L m_{\rm qs} c$, corresponding to the incoming nuclei. Using the uncertainty principle in the laboratory frame, we can bound the widths of these rapidity distribution peaks to be $\Delta \theta_{\gamma_L}\ge\hbar \gamma_L/(2D)$. In the limit of $\gamma_L\gg 1$, the energy of a quasiparticle with rapidity $\simeq\pm \theta_{\gamma_L}$ is $E\simeq\theta_{\gamma_L} c$. 

The initial stage of the heavy-ion collision in the laboratory frame can be viewed as a continuous process that transfers quasiparticles from $\theta \simeq \pm \theta_{\gamma_L}$ to the $t=0$ rapidity distribution of the quark-gluon plasma, analogous to the 1D gas Bragg pulse continuously transferring amplitude to the side rapidity peaks. It seems likely that this initial rapidity distribution contains a peak of similar width to the side peaks at rest in the laboratory frame. As in a 1D gas right after a Bragg quench, some hydrodynamization and/or local prethermalization may occur while the starting time is approached, but it is not qualitatively important.

After $t=0$, we can apply the same explanation we used for hydrodynamization in 1D gases to the evolution of the momentum distribution, which contains all the information about the relative amplitudes of the partons (quarks, antiquarks, and gluons). As in 1D gases, the momentum distribution of the partons in the quark-gluon plasma will exhibit hydrodynamization oscillations on a timescale $\tau^c_{\rm hd}= 2\pi \hbar/(\theta_{\gamma_L} c)$. $\tau^c_{\rm hd}$ depends on the mass of the quasiparticles and on $\gamma_L$. Given our assumption about the quasiparticle masses, $\tau^c_{\rm hd}$ is bounded by 0.04 (0.002) and 4 (0.2) fm/$c$ for the RHIC (LHC) experiments. If one were able to experimentally or numerically probe such oscillations, as we have done for TG gases, one could determine $m_{\rm qs}$ using $\tau^c_{\rm hd}$.

The damping of these hydrodynamization oscillations occurs within a time $\tau^d_{\rm hd}$. Notably, because of the linear dispersion in the relativistic case, this time depends on the widths of the rapidity distribution peaks and does not depend on the masses of the quasiparticles. Given our bound for the width of these peaks, we expect $\tau^d_{\rm hd}\leq 8\pi D/(3\gamma_L c)$. As in our 1D gases, we anticipate the shortest local {\it prethermalization} time for the momentum distribution of partons with high momenta to be of the same order as $\tau^d_{\rm hd}$. Although the quark-gluon plasma ultimately {\it thermalizes} locally, {\it prethermalization} is the appropriate concept for times so short that bare particles cannot traverse the interparticle separation. At longer times, there will be local thermalization and the associated decay of quasiparticles.

Assuming minimal uncertainty and using the largest values of $\gamma_L$ from each experiment, $\tau^d_{\rm hd}\simeq 1.2$ fm/$c$ for the RHIC and $\tau^d_{\rm hd}\simeq 0.05$ fm/$c$ for the LHC. These values would decrease if $\Delta \theta_{\gamma_L}$ is in fact larger than our uncertainty principle-limited lower bound, and would increase if we decrease $\gamma_L$. These times are generally in the range of or somewhat shorter than the times after which hydrodynamics starts to describe the quark-gluon plasma (recall, $0.4–1.0$ fm/$c$ for the RHIC and $0.2–0.6$ fm/$c$ for the LHC~\cite{busza_18}). Hydrodynamics cannot be used to describe dynamics until after hydrodynamization has run its course in $\sim \tau^d_{\rm hd}$. Until then, there are significant short distance spatial fluctuations and bare particle energies are still being redistributed across distant momentum modes. 

Therefore, the earliest time for which hydrodynamics can possibly work is $\sim \tau^d_{\rm hd}$. But how soon after $\tau^d_{\rm hd}$ hydrodynamics becomes a reliable description is a subtle question. That it might be roughly coincident with $\tau^d_{\rm hd}$ is supported by our study of 1D gases, where we have shown that $T^d_{\rm hd}$ scales the same as $T_{\rm dp}$, and the normalized difference between the time-evolving $f(p)$ and the local prethermalized $f(p)$ decreases from $\sim$10$\%$ to $\sim$1$\%$ between $T_{\rm dp}$ and $2T_{\rm dp}$ (see Fig.~\ref{fig:3peakDfvst}). Furthermore, in multiple studies of integrable models after quantum quenches, it has been found that observables after local prethermalization can be very close to locally thermal~\cite{calabrese_essler_review_16} (see, e.g., Ref.~\cite{Fitzpatrick11} for a systematic study of a specific family of initial states). Putting these two sets of observations together, if on the $T_{\rm dp}$ timescale $f(p)$ in the quark-gluon plasma approaches the locally prethermalized state and the locally prethermalized $f(p)$ is similar to $f(p)$ after local thermalization, hydrodynamics should be reasonably accurate soon after $\sim \tau^d_{\rm hd}$. We note that in the special case of Bragg-scattered near-integrable 1D Bose gases, generalized hydrodynamics demonstrably applies for times $\gtrsim T^d_{\rm hd}$~\cite{le2023observation}. If it turns out that the proper time $\tau^{\rm exp}_{0}$ at which hydrodynamic descriptions start to work is somewhat bigger than $\tau^d_{\rm hd}$, then there would be a minor nomenclature discrepancy, in that our usage of hydrodynamization would not coincide with the applicability of hydrodynamics, as in the original high-energy coining of the term.

As an additional note, if the collision between the heavy nuclei does not produce a quasiparticle peak about $\theta=0$, but instead simply entangles the quasiparticles with rapidities about the $\theta_{\gamma_L}$ peak with those about the $-\theta_{\gamma_L}$ peak, then hydrodynamization oscillations would not occur (as discussed for the two-peak case in Sec.~\ref{sec:2peak}). Local prethermalization would still occur with the same scaling with $\tau^d_{\rm hd}$ discussed above, assuming that all quasiparticles from one peak are entangled with all the quasiparticles from the other peak. Thus either of the two scenarios discussed in Secs.~\ref{sec:3peak} and~\ref{sec:2peak}, for three and two peaks in the rapidity distributions, respectively, are consistent with the short times for which hydrodynamics can be used to describe the experimental results. 

\section{Summary} \label{sec:summary}

Our results show that in high-energy quenches involving multimodal energy distributions of the quasi-particles, hydrodynamization and local prethermalization dynamics proceed on remarkably short timescales. These timescales are set by the amount of energy put into the system via the quench. Assuming that even nonintegrable systems have well defined (albeit typically hard to calculate) quasiparticles on such short timescales, we have qualitatively explained why hydrodynamics works so soon after the start of a heavy-ion collision.

The quasiparticle description of an interacting many-body system after a quench depends on the relevant detailed physics. However, the subsequent dynamics is rather universal. If there are three peaks in the rapidity distribution (the momentum distribution of the quasiparticles) then observables will exhibit hydrodynamization oscillations characterized by the time $T^c_{\rm hd}$. These oscillations damp out on the $T^d_{\rm hd}\propto T_{\rm dp}$ timescale, where the particular damping coefficient depends on the observable. The highest momentum components of the momentum distribution are the earliest to prethermalize, in a time $T_{\rm lp}\simeq T_{\rm dp}$. Hydrodynamization and initial local prethermalization are dominated by spatial changes on distance scales shorter than interparticle separations and by exchanges of energy among disparate energy modes. The dynamics of subsequent local prethermalization, which will generally overlap with local thermalization in nonintegrable systems, involves spatial changes that occur on distances that are of the order or greater than the interparticle separation. Energy is exchanged primarily among nearby modes and their associated prethermalization time $T_{\rm lp}$ is greater than $T_{\rm dp}$. We further showed in this work that for initial states in which there are only two peaks in the rapidity distribution, there are no hydrodynamization oscillations, even when the quench energy is the same as in a three-peak quench. In that case, the fastest prethermalization evolution still occurs in times $T_{\rm lp}\simeq T_{\rm dp}$, as in the three-peak case.

For future studies with 1D gases, we envision exploring what happens as one controllably breaks integrability, as was done, e.g., in Ref.~\cite{ben_18} for a QNC sequence. It appears particularly interesting to understand how the nature of the state at times $\sim T_{\rm dp}$ changes, as well as the interplay between local prethermalization and local thermalization, as one departs from the near-integrable regime. Beyond 1D gases, Bragg scattering quenches in unitary three-dimensional gases~\cite{eigen_2018, yin_2013} may be good candidates for studying the interplay between local prethermalization and local thermalization in far-from-integrable systems. 

\begin{acknowledgments}
We thank Sarang Gopalakrishnan for discussions early in this work. Y.Z.~acknowledges support from the Dodge Family Postdoc Fellowship at the University of Oklahoma. M.R.~acknowledges support from the National Science Foundation under Grant No.~PHY-2309146. D.S.W.~and Y.L. acknowledge support from the National Science Foundation under Grants No.~PHY-2012039 and PHY-2409213. 
\end{acknowledgments}

\appendix

\section{Experiments and theory with\\ Bragg pulses}\label{app:exp-the-bra}

\subsection{Experimental setup}

The experiments in Ref.~\cite{le2023observation} and the experiments in this work start with a Bose-Einstein condensate of $^{87}$Rb atoms in the $F=1$, $m_F=1$ ground state confined in a red-detuned crossed dipole trap. A bundle of near-zero temperature 1D gases confined in separate ``tubes'' is created by adiabatically ramping up a blue-detuned 2D lattice to 40$E_r$~\cite{kinoshita2005all}, where $E_r=\hbar^2k_0^2/(2m)$ is the recoil energy, $k_0=2\pi/775$ nm is the lattice wave vector, and $m$ is the mass of a $^{87}$Rb atom. Theoretically, each individual tube can be modeled by adding a confining potential $U(z)$ to the Lieb-Liniger Hamiltonian~\cite{lieb1963exact},
\begin{equation}\label{eq:H_lieb_liniger}
{\cal H}_\text{LL}=\sum_{j=1}^{N}\left[-\frac{\hbar^2}{2m}\frac{\partial^2}{\partial z^2_j}+U(z_j)\right]+g\sum_{1\leq j < l \leq N}\delta(z_j-z_l) \,,
\end{equation}
where $N$ is the number of atoms in the tube, and $g$ is the strength of contact interaction, which depends on the depth of the 2D lattice~\cite{olshanii1998atomic}. $U(z)$, generated by the crossed dipole trap, is modeled as a Gaussian-shaped potential,
\begin{equation}\label{eq:U_ini}
U(z)=U_0\left[1-\exp\left(-\frac{2z^2}{w^2}\right)\right],
\end{equation}
where $U_0$ is the depth of the Gaussian trap and $w$ is its waist. The Lieb-Liniger Hamiltonian [Eq.~(\ref{eq:H_lieb_liniger}) with $U(z_j)=0$] is exactly solvable via the Bethe ansatz~\cite{lieb1963exact, yang1969thermodynamics}. Observables in the ground state only depend on the dimensionless parameter $\gamma=mg/(\rho\hbar^2)$, where $\rho$ is the 1D atom density. For the trapped case, within the local density approximation, one can define the local $\gamma(z)=mg/[\rho(z)\hbar^2]$ and use the average value of $\gamma$, $\bar \gamma = \int dz \rho(z) \gamma(z) / [\int dz \rho(z)]$, to characterize the system. Specifically, we use the weighted average $\bar \gamma_0$ of $\bar \gamma$ in the initial state prior to the quenches in the array of 1D gases.

\subsection{Tonks-Girardeau theory}

We carry out our theoretical analyses in the strong coupling Tonks–Girardeau (TG) limit ($\gamma\to\infty$) of Hamiltonian~(\ref{eq:H_lieb_liniger}), where out-of-equilibrium momentum and rapidity distributions can be readily calculated for arbitrary initial states. As shown in Ref.~\cite{le2023observation} and in Appendix~\ref{app:nsbp}, the theoretical results in that limit parallel the experimental observations of hydrodynamization and local prethermalization at finite $\gamma$. In our calculations, we use the low-density regime of the lattice hard-core boson Hamiltonian,
\begin{equation}\label{eq:HCB}
{\cal H}_{\rm HCB}=-J\sum_{j=1}^{N_s-1}\big(\hat b^\dagger_{j+1}\hat b^{}_j + {\rm H.c.}\big)+\sum_{j=1}^{N_s}U(z_j)\hat b^\dagger_j\hat b^{}_j\,,
\end{equation}
where $J$ is the hopping amplitude of a hard-core boson, and $N_s$ is the total number of lattice sites. The boson creation (annihilation) operator $\hat b^\dagger_j$ ($\hat b_j$) creates (annihilates) a boson at lattice site $j$, and the hard-core constraint is enforced by $\hat b_j^2=\hat b^{\dagger2}_j=0$. The position of site $j$ on the lattice is set to be $z_j=(j-N_s/2)a$, with $a$ the lattice spacing. In the limit in which the site occupation $n_j=\langle\hat {b}^{\dagger}_j\hat {b}^{}_j\rangle\to0$ on all sites, the lattice hard-core boson Hamiltonian~(\ref{eq:HCB}) is equivalent to the continuum Hamiltonian~(\ref{eq:H_lieb_liniger}) in the TG limit~\cite{rigol_muramatsu_05}. The parameters of these two Hamiltonians are connected by $J=\hbar^2/(2ma^2)$. 

The main experimental observable is the momentum distribution $f(p)$, which is measured via a time-of-flight expansion~\cite{bloch_08} after the 1D interactions have been essentially turned off~\cite{wilson_malvania_20}. The redistribution of energy among momentum modes is the probed by studying the integrated kinetic energy per particle
\begin{equation}\label{eq:barEp}
E_{\bar p}=\int_{\bar p-\Delta_p/2}^{\bar p+\Delta_p/2}dp\,\frac{f(p)p^2}{2m},
\end{equation}
within the momentum intervals $\Delta_p$ centered around $\bar p$. For the TG theory, $f(p)$ is written as (up to a normalization constant)
\begin{equation}\label{eq:momentum}
    f(p) \propto \langle \Psi_B| \hat b^\dagger_p \hat b^{}_p |\Psi_B\rangle = \sum_{jl} e^{ip(z_j-z_l)/\hbar} \langle \Psi_B| \hat b^\dagger_{j} \hat b^{}_{l} |\Psi_B\rangle,
\end{equation}
where $\hat b_p^{\dagger}$ ($\hat b_p$) creates (annihilates) a hard-core boson with momentum $p$, and $|\Psi_B\rangle$ is the boson wave function. The in- and out-of-equilibrium equal-time correlation function $\langle \Psi_B| \hat b^\dagger_{j} \hat b^{}_{l} |\Psi_B\rangle$ can be efficiently computed by mapping the hard-core bosons onto noninteracting fermions via a Jordan-Wigner transformation \cite{jordan_wigner_28}, 
\begin{equation}
    \hat b^\dagger_j=\hat c^\dagger_j e^{-i\pi\sum_{l<j}\hat c^\dagger_l\hat c_l}\,,
\end{equation}
where $\hat c^{\dagger}_j$ ($\hat c_j$) is the fermion creation (annihilation) operator at site $j$, and then using properties of Slater determinants~\cite{rigol_muramatsu_05, rigol_muramatsu_05c}. In the TG limit, the quasiparticles are the noninteracting fermions onto which the hard-core bosons are mapped. Therefore, the rapidity distribution $f(\theta)$ is computed to be
\begin{equation}
f(\theta) \propto \langle \Psi_F| \hat c^\dagger_\theta \hat c^{}_\theta |\Psi_F\rangle = \sum_{jl} e^{i\theta(z_j-z_l)/\hbar} \langle \Psi_F| \hat c^\dagger_{j} \hat c^{}_{l} |\Psi_F\rangle,
\end{equation}
where $\hat c_\theta^{\dagger}$ ($\hat c^{}_\theta$) creates (annihilates) a noninteracting fermion with rapidity $\theta$, and $|\Psi_F\rangle$ is the fermionic wave function.

\subsection{Bragg pulses}\label{sec:bragg}

To create out-of-equilibrium states that feature hydrodynamization, in Ref.~\cite{le2023observation} and in Appendix~\ref{app:nsbp}, we apply a Bragg pulse quench to the system with a $\lambda$ wavelength standing wave. That is, we turned on an additional potential, 
\begin{equation}\label{eq:U_pulse}
U_{\rm pulse}(z)=U_{\rm pulse}\cos^2(kz)\,,
\end{equation}
with wave number $k$ and amplitude $U_{\rm pulse}$, for a short period of time $t_{\rm pulse}$. In our experiments $k\approx k_0$, where $k_0$ is the wave number of the 2D lattice. After the pulse, the system was allowed to evolve for different times $t_{\rm ev}$ before the momentum distribution $f(p)$ was measured. 

The effect of the Bragg pulse can be understood analytically in the Raman-Nath regime ($t_{\rm pulse}\to0$) for the homogeneous TG gas. The pulse in this regime, known as a Kapitza-Dirac pulse, ``kicks'' the particles and quasiparticles before they have had time to move. Such a kick generates a state in which the quasiparticles are in a superposition of plane waves that differ by even integer multiples of the Bragg momentum~\cite{caux2016separation}. The many-quasiparticle (noninteracting fermion) wave function changes from that of a Fermi sea $|\Psi_0\rangle=\prod_{|\theta|<\theta_F}\hat c^{\dagger}_\theta|0\rangle$ to
\begin{equation}\label{eq:braggwf}
|\Psi(t_{\rm ev})\rangle\!=\!\!\!\!\prod_{|\theta|<\theta_F}\!\sum_n I_{n}(-ib)e^{-it_{\rm ev}(\theta+n2\hbar k)^2/2m\hbar}\hat c^{\dagger}_{\theta+n2\hbar k}|0\rangle.
\end{equation}
$I_{n}(-ib)$ are the modified Bessel functions of the first kind, where the integer $n$ is the order of the Bragg peak and $b$ depends on the area of the Bragg pulse. In a homogeneous system, the rapidity distribution $f(\theta)$ is conserved after the quench, but the momentum distribution $f(p)$ for the bare particles evolves in time. The time evolution of the relative phases of different $\hat c^\dagger_\theta$ enters $f(p)$ via the Bose-Fermi mapping, so $f(p)$ changes in time despite $f(\theta)$ being constant. That is, although there is no redistribution of energy among the quasiparticles, there is a redistribution of energy among different momentum modes of the particles. 

\section{A new single Bragg pulse}\label{app:nsbp}

In the experimental results reported in this work, we use different pulses and pulse sequences to explore the hydrodynamization process and its absence under different conditions. Our single Bragg pulse has $t_{\rm pulse}=10$ $\mu$s and $U_{\rm pulse}=18$ $E_r$, and it is applied to a system that is initially in a trap with $U_0=2.6$ $E_r$ and $\bar\gamma_0=3.4$. We choose these Bragg pulse parameters to correspond to the second Bragg pulse of the Newton's cradle sequence described in Sec.~\ref{sec:qnc}, which allows us to most directly compare the evolution of two- and three-peak experimental quenches.  

\begin{figure}[!b]
    \includegraphics[width=0.985\columnwidth]{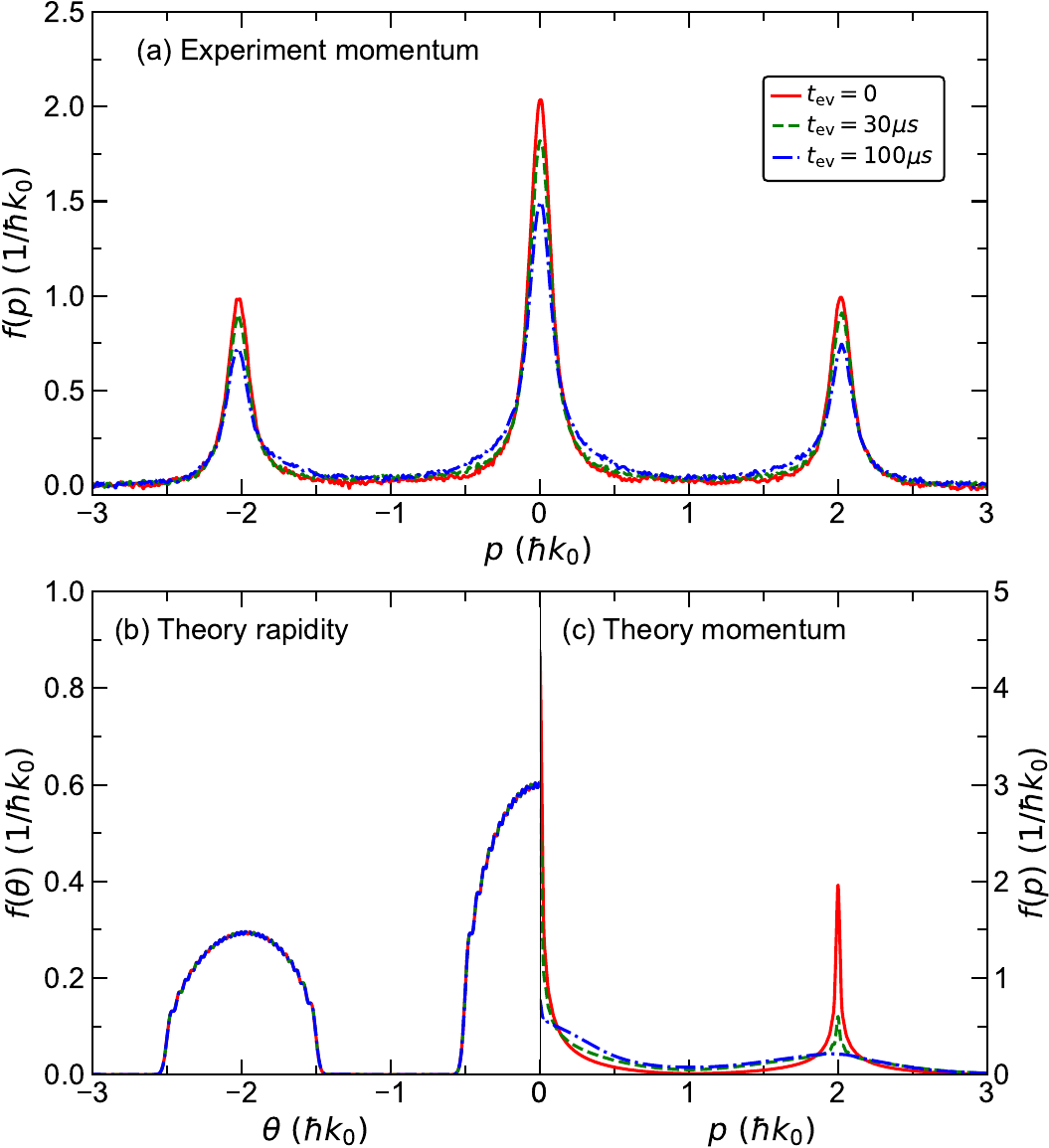}
    \vspace{-0.15cm}
    \caption{{\it Time evolution of the momentum and rapidity distributions after a 10 $\mu$s Bragg pulse.} (a) Experimental results for the momentum distribution for $\bar \gamma_0=3.4$. (b), (c) Corresponding theoretical results for the rapidity (b) and momentum (c) distribution of a single 1D gas in the TG limit. In the theory, the trap parameters and the average energy per atom are the same as in the experiment before the quench.}
    \label{fig:profile10mus}
\end{figure}

In Fig.~\ref{fig:profile10mus}(a), we plot our experimental results for the momentum distribution right after the Bragg pulse ($t_{\rm ev}=0$), and at two later times. The most salient observation from these curves is that the momentum peaks broaden with time. In Figs.~\ref{fig:profile10mus}(b) and~\ref{fig:profile10mus}(c), we show theoretical results for the rapidity and momentum distributions, respectively, following the same Bragg pulse. The theory is for a single 1D gas in the TG limit with the same trap parameters as in the experiment. We set the atom number to be $N=29$, so that the energy per particle in the ground state of the trapped TG gas before the pulse is the same as in the experimental system for the bundle of tubes. The rapidity distribution [Fig.~\ref{fig:profile10mus}(b)] does not evolve significantly during the time in which the momentum distribution [Fig.~\ref{fig:profile10mus}(c)] evolves. Since the experimental system has an array of tubes with different atom numbers and finite $\gamma$, we do not expect our theoretical results for a single tube with $\gamma=\infty$ (initially in the ground state) to quantitatively match the experiment~\cite{malvania_zhang_21}. Our goal is to espouse a simple theoretical description that allows us to qualitatively describe and understand the experimentally observed phenomena.

\begin{figure}[!t]
    \includegraphics[width=0.985\columnwidth]{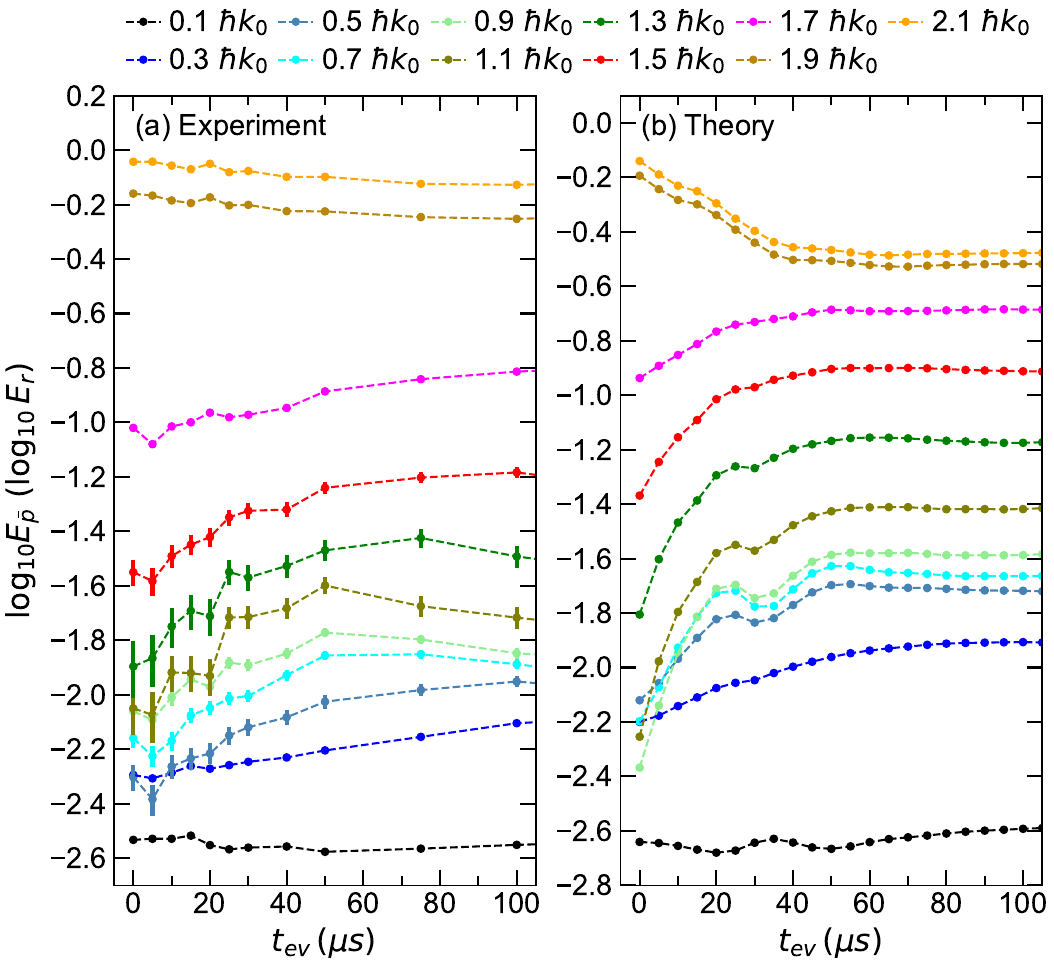}
    \vspace{-0.15cm}
    \caption{{\it Hydrodynamization after a 10 $\mu$s Bragg pulse quench.} Each curve shows the time evolution of the integrated energy $E_{\bar p}$ over a $0.2\hbar k_0$ wide momentum group [see Eq.~\eqref{eq:barEp}]. (a) Experimental results for $\bar \gamma_0=3.4$. (b) Theory calculation for a single 1D gas in the TG limit with the same trap parameters and average energy per atom as in the experiment before the quench. The dashed lines are guides to the eye.}
    \label{fig:exp10mus}
\end{figure}

In Fig.~\ref{fig:exp10mus}(a), we plot our experimental results for $E_{\bar p}$ [see Eq.~\eqref{eq:barEp}] as a function of $t_{\rm ev}$ for momentum intervals $\Delta_p=0.2 \hbar k_0$. As in the results in Ref.~\cite{le2023observation}, for which a shorter but stronger Bragg pulse was used, the $E_{\bar p}$ curves exhibit two clear features on the hydrodynamization coherence timescale $T^c_{\rm hd}=66$ $\mu$s [see Eq.~\eqref{eq:thd}]. First, the curves with intermediate $\bar p$ between the $n=0$ and $\pm 1$ Bragg peaks ($\bar p$ from $0.5\hbar k_0$ to $1.5\hbar k_0$) rise rapidly. Second, there are oscillations with a period that is close $T^c_{\rm hd}/2$; these are most clearly seen in the curves with $\bar p$ close to $0$ or $2\hbar k_0$. The kinetic energy curves in Fig.~\ref{fig:exp10mus}(a) are qualitatively similar to those in Ref.~\cite{le2023observation}, illustrating that the hydrodynamization process is robust. Small changes arise mostly because the difference in the experimental parameters modify $T_{\rm dp}$ (when $\bar \gamma_0$ is greater, $\theta_F$ is smaller).

In Fig.~\ref{fig:exp10mus}(b), we plot results from our numerical simulations of a single 1D gas in the TG limit with the same trap parameters as in the experiment. The theoretical $E_{\bar p}$ curves capture the main experimental features. One can see both the rapid increase in the energies of the modes near $\bar p\sim \hbar k_0$, and the oscillations in energies of the modes with $\bar p$ close to $0$ or $2\hbar k_0$. Evidently, averaging over an array of 1D gases does not change the qualitative behavior of what occurs in a single tube. We note that the changes in the experimental curves are about half as large as those in the theoretical curves, possibly a result of averaging and possibly simply because the relationships between rapidities and momenta depend on the strength $g$ of the contact interaction.

\section{Insensitivity to the trap and\\ the Bragg pulse}\label{app:initialstate}

Equation~(\ref{eq:braggwf}) assumes that the state before the Bragg pulse is the ground state of a homogeneous TG gas, and that the Bragg pulse is infinitely short. In our experiments and their modeling, the initial state is inhomogeneous, because of the trap needed to confine the atoms, and the pulse has a finite duration, because its power is finite. In this appendix, we explore theoretically the effects of inhomogeneity and the Bragg pulse used on the dynamics of a single TG gas.

\begin{figure}[!t]
    \includegraphics[width=0.985\columnwidth]{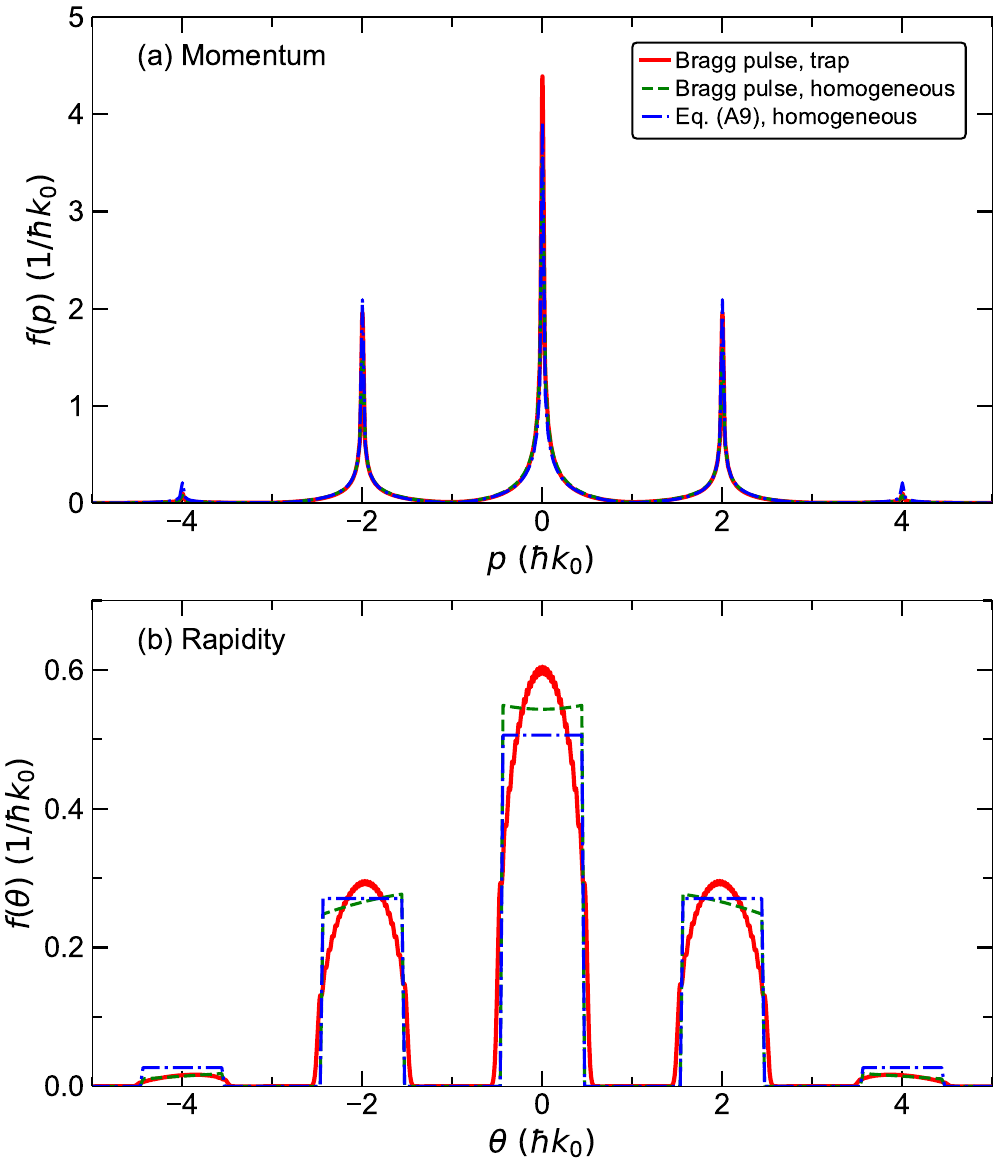}
    \vspace{-0.15cm}
    \caption{{\it Theoretical momentum and rapidity distributions at $t_{\rm ev}=0$.} (a) $f(p)$ and (b) $f(\theta)$ for a single TG gas. We show results for a trapped gas after a 10 $\mu$s Bragg pulse (continuous line), a homogeneous gas after a 10 $\mu$s Bragg pulse (dashed line), and a homogeneous gas after a Kapitza-Dirac pulse [as per Eq.~\eqref{eq:braggwf}] with the same area under the pulse as the other two cases (dashed-dotted line). In all three cases, the number of particles is $N=29$, and the energy per particle is the same before the pulse (see text).}
    \label{fig:profilecompare}
\end{figure}

In order to isolate the effects of inhomogeneity, we study the case in which the initial state is the ground state of a homogeneous TG gas with $N=29$ particles in a 1D system with periodic boundary conditions [we set $U(z)=0$], and apply a $10$ $\mu$s Bragg scattering pulse [Eq.~(\ref{eq:U_pulse})]. We select the same number of particles and apply the same Bragg pulse as in Appendix~\ref{app:nsbp}. To most cleanly compare the trapped and homogeneous cases, we choose the size of the homogeneous system so that the energy per particle before the Bragg pulse is the same as in the trapped case. The momentum and rapidity distributions obtained right after the Bragg pulse, for both the trapped and homogeneous case, are plotted in Figs.~\ref{fig:profilecompare}(a) and~\ref{fig:profilecompare}(b), respectively. Away from $p=\pm 2nk_0$, for which small differences are seen in Fig.~\ref{fig:profilecompare}(a), the momentum distributions of the trapped and the homogeneous systems are indistinguishable from each other. The corresponding rapidity distributions in Fig.~\ref{fig:profilecompare}(b) are also very close to each other, only differing, as expected, by the fact that in the homogeneous case there are Fermi sea-like discontinuities at the edges of each peak.

To isolate the effects of the finite duration of our Bragg pulse, in Figs.~\ref{fig:profilecompare}(a) and~\ref{fig:profilecompare}(b) we also plot the momentum and rapidity distributions, respectively, for the state predicted by Eq.~\eqref{eq:braggwf} for a Kapitza-Dirac pulse with the same area as our Bragg pulse applied to the ground state of the homogeneous case. Figure~\ref{fig:profilecompare}(b) shows that the main effect of the finite duration of the Bragg pulse is that it reduces the transfer of quasiparticles out of the initial Fermi sea, resulting in a slightly lower population of its $n=\pm1,\pm2$ ``replicas''. Figure~\ref{fig:profilecompare}(a) shows that this results in slightly lower $n=\pm1,\pm2$ peaks in the momentum distribution after the Bragg pulse. Away from $p=\pm 2nk_0$, the momentum distributions are all indistinguishable from each other.

\begin{figure}[!t]
    \includegraphics[width=.985\columnwidth]{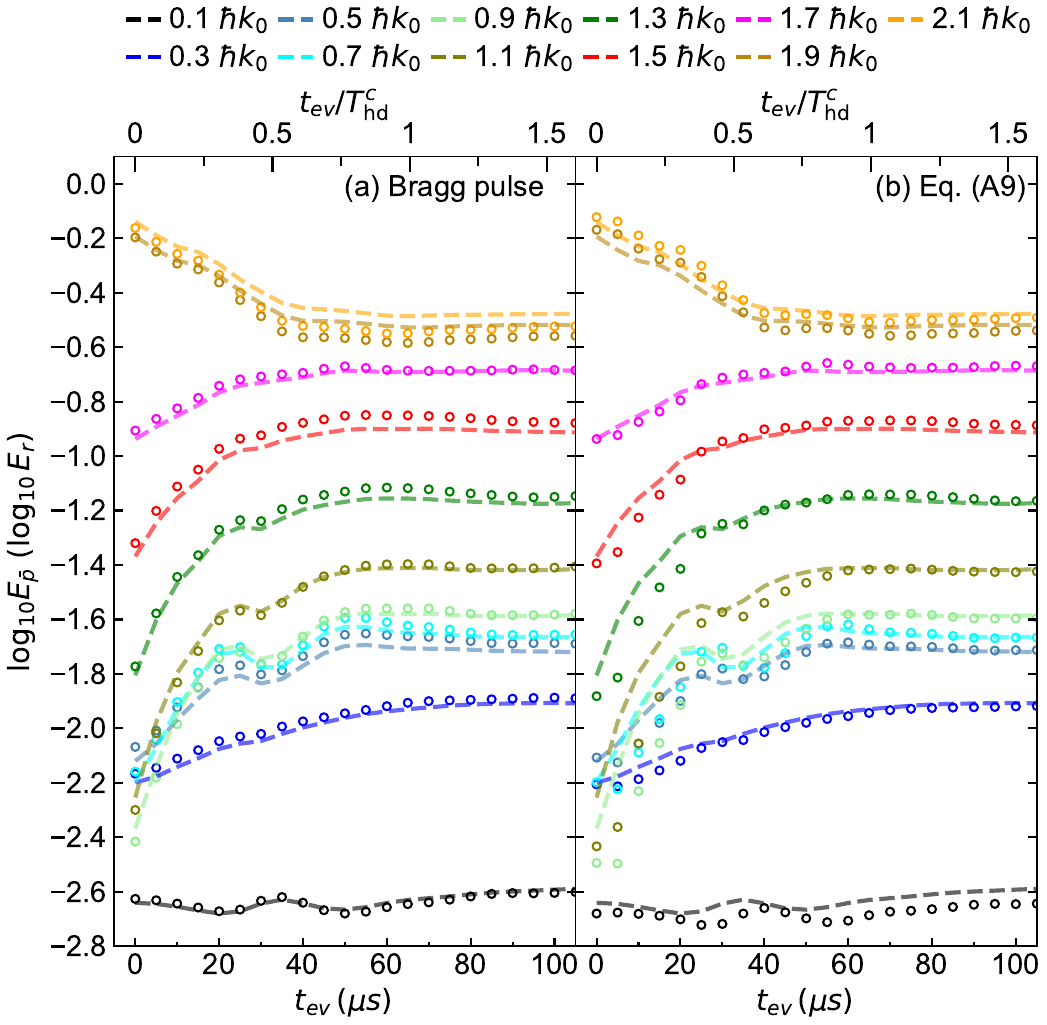}
    \vspace{-0.15cm}
    \caption{{\it Time evolution of $E_{\bar p}$ for the homogeneous TG gas.} The open symbols show the results after: (a) a 10 $\mu$s Bragg pulse, and (b) a Kapitza-Dirac pulse [as per Eq.~\eqref{eq:braggwf}] with the same area under the pulse as in (a). The dashed lines show the results for the trapped TG gas reported in Fig.~\ref{fig:exp10mus}(b). The number of particles, $N=29$, and the energy per particle before the pulse are the same for the homogeneous TG gas as for the trapped TG gas. To obtain each curve, we integrated within a momentum interval of 0.2 $\hbar k_0$. The upper $x$ axis shows the dimensionless quantity $t_{\rm ev}/T^c_{\rm hd}$ [see Eq.~\eqref{eq:thd}].}
    \label{fig:Ekcompare}
\end{figure}

In Fig.~\ref{fig:Ekcompare}, we show the evolution of the integrated kinetic energy $E_{\bar p}$ after the Bragg pulse [Fig.~\ref{fig:Ekcompare}(a)] and after the Kapitza-Dirac pulse [Fig.~\ref{fig:Ekcompare}(b)] for the homogeneous TG gas (empty symbols). Since the two are very similar to each other, it is clear that the finite duration of the pulse, for which there can be many-body evolution during the pulse, does not in fact introduce qualitative changes. In Figs.~\ref{fig:Ekcompare}(a) and~\ref{fig:Ekcompare}(b) we also show, as thin dashed lines, the results for the trapped system already reported in Fig.~\ref{fig:exp10mus}(b). The evolution in the homogeneous and trapped cases are very similar. In all cases there is a rapid redistribution of kinetic energy, and oscillations of $E_{\bar p}$, in a time $T^c_{\rm hd}/2$. Hence, hydrodynamization, as seen in the experiments and in their corresponding modeling using a trapped TG gas, is not qualitatively affected by either the fact that the system is inhomogeneous or by the finite duration of the Bragg pulse. Therefore, modeling that directly uses Eq.~\eqref{eq:braggwf}, with the appropriate initial state and area of the Bragg pulse, provides a quantitatively reliable description of the dynamics of the trapped TG gas following the finite-duration Bragg pulse.

Next, we explore the local prethermalization of the momentum distribution in the presence and absence of a trap. Local prethermalization refers to the equilibration of the local momentum distribution to the GGE prediction~\cite{rigol2007relaxation, ilievski_denardis_15, vidmar2016generalized, calabrese_essler_review_16}. During local prethermalization, the overall density distribution does not change significantly. After local prethermalization, trapped systems exhibit global dynamics during which the density distribution changes and with it the local and global rapidity and momentum distributions. Such a long-wavelength and long-time dynamics in an integrable system is described by generalized hydrodynamics~\cite{castro2016emergent, bertini2016transport, doyon_20, Alba_2021, Bastianello_2022}, as shown in recent experiments~\cite{schemmer2019generalized, malvania_zhang_21, joerg_21, joerg_22, yang2023phantom}.  

\begin{figure}[!t]
    \includegraphics[width=0.95\columnwidth]{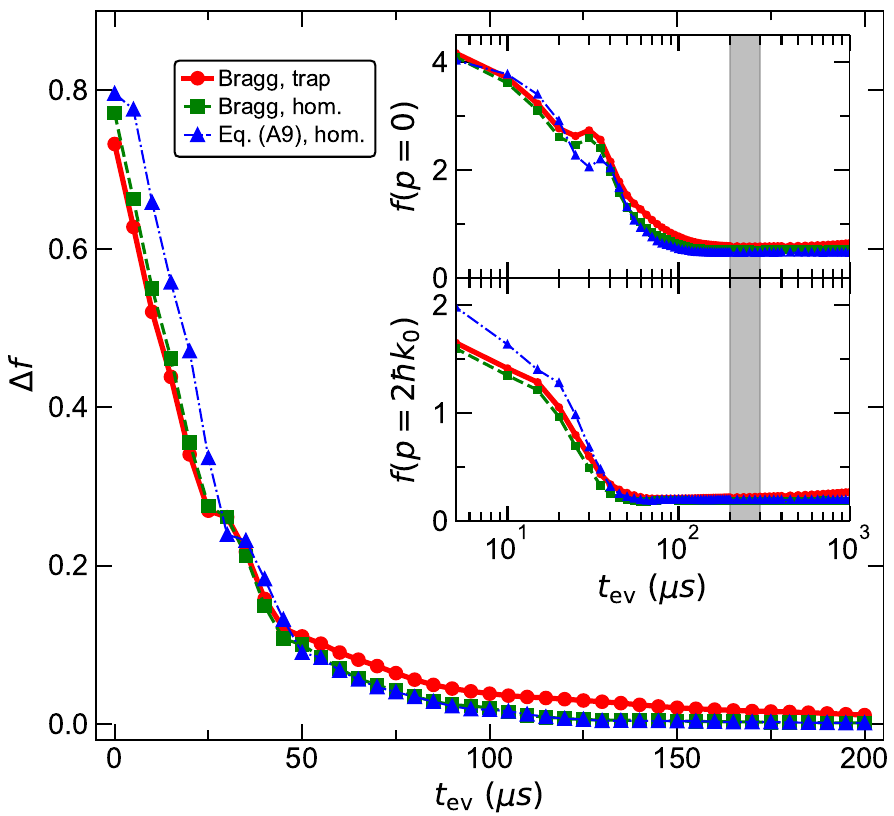}
    \vspace{-0.15cm}
    \caption{{\it Time evolution of the momentum distributions of trapped and homogeneous TG gases.}  Time evolution of $\Delta f$ [see Eq.~\eqref{eq:diffDdef}] for a trapped gas after a 10 $\mu$s Bragg pulse (circles), a homogeneous gas after a 10 $\mu$s Bragg pulse (squares), and a homogeneous gas after a Kapitza-Dirac pulse [as per Eq.~\eqref{eq:braggwf}] with the same area under the pulse as the other two cases (triangles). In all three cases, the number of particles is $N=29$, and the energy per particle is the same before the pulse. (Insets) Time evolution of $f(p=0)$ (upper inset) and  $f(p=2\hbar k_0)$ (lower inset) for the same systems considered in the main panel. The vertical gray bands highlight the time interval used to calculate the time-averaged $\bar f(p)$ input in the calculation of $\Delta f(t_{\rm ev})$.}
    \label{fig:Dfcompare}
\end{figure}

The local prethermalization time for $f(p)$ is in general momentum-dependent, with the occupation of lower momenta (determined by correlations over longer distances) requiring longer prethermalization times. We discuss this dependence in detail in Sec.~\ref{sec:3peak}. The insets in Fig.~\ref{fig:Dfcompare} show the time evolution of $f(p=0)$ (top inset) and $f(p=2\hbar k_0)$ (bottom inset), both following the Bragg pulse in the trapped (red points) and homogeneous systems (green points) and after the Kapitza-Dirac pulse in the homogeneous system (blue points). The initial changes are qualitatively the same for the different pulses. As expected, $f(p=0)$ takes longer to prethermalize. After first equilibrating to a nearly time-independent value, $f(p=0)$ and $f(p=2\hbar k_0)$ in the trapped system depart from that value at $t\sim 1$ ms because the density distribution begins to change in the trap. In the homogeneous cases the density distribution does not change and $f(p=0)$ remains unchanged (except for the inevitable revivals that ultimately occur in finite systems). 

In the insets in Fig.~\ref{fig:Dfcompare} one can see that the values of $f(p=0)$ and $f(p=2\hbar k_0)$ for the three systems considered are nearly time-independent in the interval $t_{\rm ev} \in [200,300]\,\mu$s highlighted by the gray band. We use six times spaced by $\delta t_{\rm ev}=20 \mu s$ in that interval to calculate the time-averaged momentum distribution $\bar f(p)$. We take $\bar f(p)$ to be the locally equilibrated momentum distribution. To characterize the local prethermalization of the momentum distribution, we calculate the relative difference $\Delta f(t_{\rm ev})$ [see Eq.~\eqref{eq:diffDdef}]. The main panel in Fig.~\ref{fig:Dfcompare} shows the time evolution of $\Delta f$ for the trapped TG gas after a 10 $\mu$s Bragg pulse, and for the homogeneous TG gas after a 10 $\mu$s Bragg pulse and after a Kapitza-Dirac pulse. The three curves are very similar to each other, differing only at the earliest time, presumably because the Bragg pulse calculations get a local prethermalization head start during the pulse. The delay seems unimportant for the characterization of the local prethermalization of $f(p)$. As with hydrodynamization, local prethermalization is essentially the same in the presence and absence of a trap, as well as with and without pulses with a finite-time duration.

Having shown that there is no need to introduce the trapping potential or to carry out a finite-duration pulse to reproduce the experimental observations, in the main text we take a final step in simplifying our theoretical modeling. In Eq.~\eqref{eq:braggwf}, the quasiparticles are in infinite superpositions of plane waves, corresponding to the infinite Bragg peaks whose weights are determined by the area under the pulse via the Bessel function. In the main text, we consider the cases in which only three or two Bragg peaks are present, to avoid the small complications associated with the small higher order peaks.

\section{Three-peak states: \\ Density distribution and $p_{50}$}\label{app:nopulse_3p}

\begin{figure}[!t]
    \includegraphics[width=0.985\columnwidth]{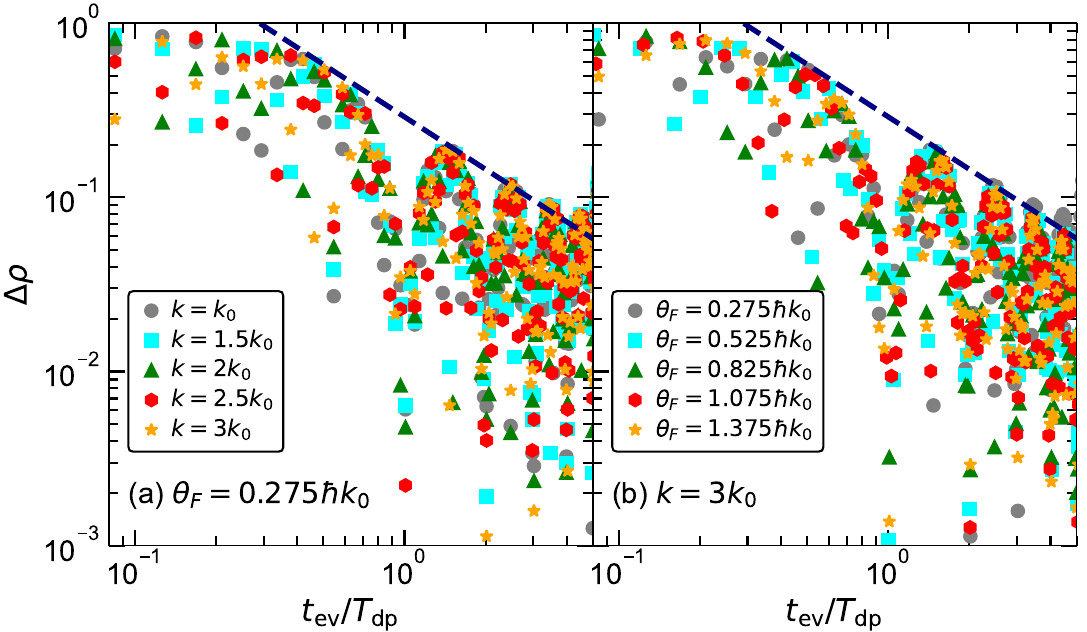}
    \vspace{-0.15cm}
    \caption{{\it Prethermalization of $\rho(z)$ for three-peak states.} Same as Fig.~\ref{fig:3peakDfvst} but for $\Delta \rho$ [see Eq.~\eqref{eq:diffRdef}]. The dashed lines, which are guides to the eye, show $(t_{\rm ev}/T_{\rm dp})^{-1}$ behavior.}
    \label{fig:3peakDrvst}
\end{figure}

In contrast to the momentum distribution, we do not expect the TG gas to serve as a model for the quantum dynamics of the density distribution in the experiments. That is because the matrix elements of few-body observables in noninteracting models are qualitatively different from those in interacting models~\cite{zhang_vidmar_22}. The density dynamics of a TG gas and a noninteracting Fermi gas are identical, while there is no such mapping for a Lieb-Liniger gas away from the TG limit. The Gaussian nature of the energy eigenstates in noninteracting models makes the observables equilibrate polynomially in time~\cite{Cramer_2008, Gluza_2016, murthy19, Gluza_2019, Patrycja_23}, as opposed to exponentially.

In Fig.~\ref{fig:3peakDrvst}, we plot the time evolution of the normalized integrated difference between the density distribution of our TG gases at different evolution times and the constant density $N/L$ achieved after prethermalization
\begin{equation}\label{eq:diffRdef}
    \Delta \rho(t_{\rm ev})= \frac{1}{N}\int dz \left|\rho(z,t_{\rm ev})- \frac NL\right|
\end{equation}
vs $t_{\rm ev}/T_{\rm dp}$. Results for $\Delta \rho(t_{\rm ev})$ are shown for systems with the same number of particles ($N=11$) and different $k$ [Fig.~\ref{fig:3peakDrvst}(a)], and for systems with the same value of $k=3k_0$ and different $N$ [Fig.~\ref{fig:3peakDrvst}(b)], which parallel the results for $\Delta f(t_{\rm ev})$ in Figs.~\ref{fig:3peakDfvst}(a) and~\ref{fig:3peakDfvst}(b), respectively. In contrast to $\Delta f(t_{\rm ev})$, $\Delta \rho(t_{\rm ev})$ exhibits large fast changes on the hydrodynamization timescale, the result of the oscillations of $\rho(z,t_{\rm ev})$ discussed earlier (and shown in Fig.~1(c) of Ref.~\cite{le2023observation}). There are also oscillations with the longer $T_{\rm dp}$ period. Most notably and as expected, the amplitude of the oscillations decays polynomially (as opposed to exponentially) in time as $(t_{\rm ev}/T_{\rm dp})^{-1}$, as demarcated by the dashed lines in the plots (see also Ref.~\cite{zhang_vidmar_21}). We expect this decay to change into an exponential decay as one moves away from the TG limit, a change that could be explored in future studies.

\begin{figure}[!t]
    \includegraphics[width=0.985\columnwidth]{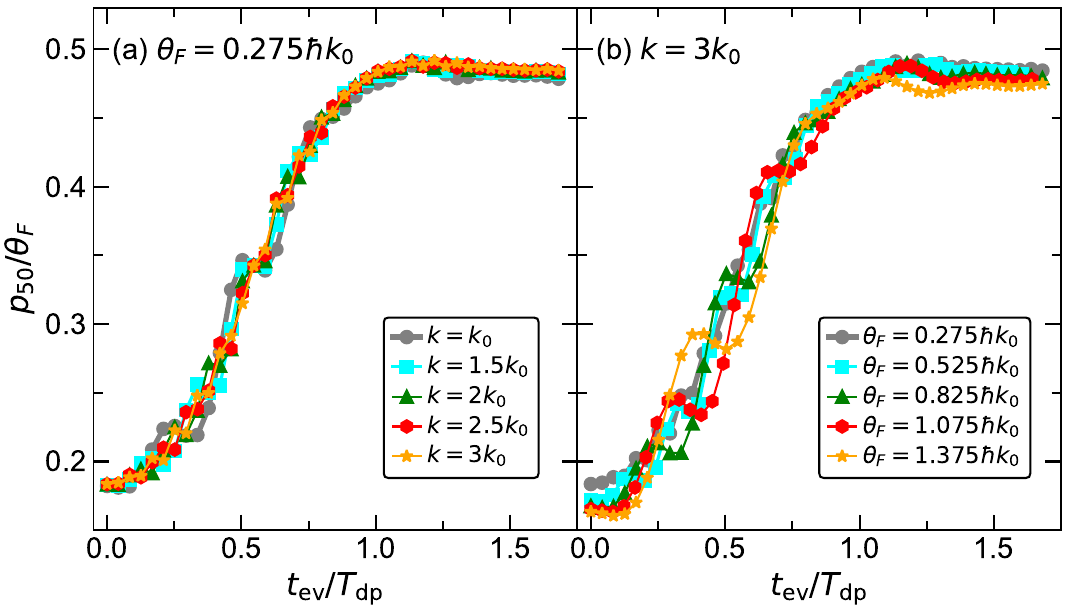}
    \vspace{-0.15cm}
    \caption{{\it Prethermalization of $p_{50}$ for three-peak states.} $p_{50}/\theta_F$ vs the dimensionless time $t_{\rm ev}/T_{\rm dp}$. (a) We fix $N=11$ ($\theta_F=0.275\hbar k_0$) and show results for different $k$. (b) We fix $k=3k_0$ and show results for different $N$ ($\theta_F$).}
    \label{fig:p50vst}
\end{figure}

We close our discussion of the three-peak case by connecting our results for  prethermalization in this work to the local prethermalization results reported in Ref.~\cite{le2023observation}. There we studied the time evolution of ``$p_{\rm f}$,'' which is defined so that $\pm p_{\rm f}$ are the momenta enclosing $f\%$ of the atoms in the central peak. Since the results were found to be qualitatively similar for $f$ in the range of 40 to 60, we focused on $f=50$, 
\begin{equation}\label{eq:p50}
\int_{-p_{\rm 50}}^{p_{\rm 50}}f(p)dp = \frac{A}{2},
\end{equation}
where $A$ is the area of the central peak.

In Fig.~\ref{fig:p50vst}, we plot the theoretical time evolution of $p_{50}/\theta_F$ vs $t_{\rm ev}/T_{\rm dp}$ for three-peak states with the same number of particles ($N=11$) when we change $k$ [Fig.~\ref{fig:p50vst}(a)], and with the same value of $k=3k_0$ when we change the number of particles [Fig.~\ref{fig:p50vst}(b)]. The curves collapse onto each other. That is, the prethermalization time for $p_{50}$ is proportional $T_{\rm dp}$. It is remarkable that the same holds true for the experimental results reported in Ref.~\cite{le2023observation}, for which the contact interaction strength $g$ had a finite constant value and the 1D density was varied to change the width of the rapidity distribution. While the local prethermalization time was found to be about three times longer in the experiments than in the TG theory, its scaling with $T_{\rm dp}$ was the same. Those results suggest that the proportionality constant between the local prethermalization time and $T_{\rm dp}$ depends on $g$ [see Eq.~\eqref{eq:H_lieb_liniger}] and not on the density (or $\theta_F$). This is unlike the thermal equilibrium properties of the Lieb-Liniger gas, which are fully determined by $\gamma$.

\section{Two-peak states: Density distribution}\label{sec:nopulse_2p}

\begin{figure}[!h]
\includegraphics[width=0.985\columnwidth]{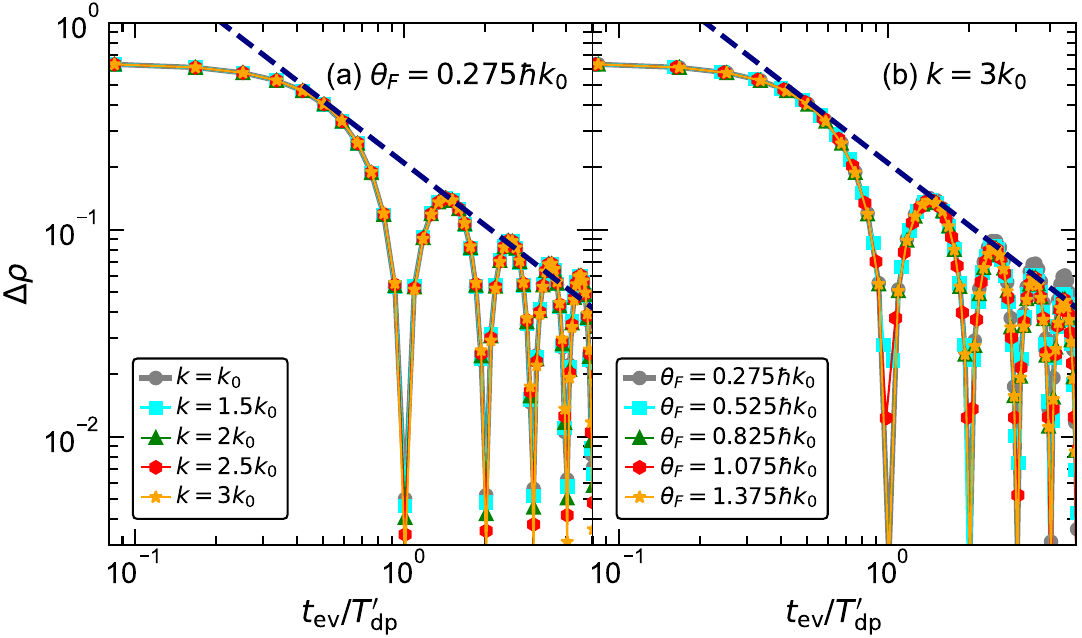}
    \vspace{-0.15cm}
    \caption{{\it Local prethermalization of $\rho(z)$ for the ``two-peak'' state.} Same as Fig.~\ref{fig:2peakDfvst} but for $\Delta \rho$ [see Eq.~\eqref{eq:diffRdef}]. The dashed lines show $(t_{\rm ev}/T'_{\rm dp})^{-1}$ behavior as in Fig.~\ref{fig:3peakDrvst}.}
    \label{fig:2peakDrvst}
\end{figure}

In Fig.~\ref{fig:2peakDrvst} we report our results for two-peak states for $\Delta \rho$ [see Eq.~\eqref{eq:diffRdef}] vs $t_{\rm ev}/T_{\rm dp}$. They parallel the results for the three-peak states shown in Fig.~\ref{fig:3peakDrvst}. All the features discussed related to the prethermalization of the density distribution of the three-peak states are apparent in the two-peak case. The main difference is that (as expected from our discussion in the main text) the hydrodynamization oscillations of the density that can be seen in Fig.~\ref{fig:3peakDrvst} are not present in Fig.~\ref{fig:2peakDrvst}. The only oscillations in Fig.~\ref{fig:2peakDrvst} are the ones with a period $T'_{\rm dp}$.

\bibliography{Reference}

\end{document}